\documentclass[fleqn,usenatbib]{mnras}

\usepackage{newtxtext,newtxmath}
\usepackage{color}

\usepackage[T1]{fontenc}
\usepackage{ae,aecompl}

\usepackage{graphicx}	
\usepackage{amsmath}	
\usepackage{amssymb}	
\usepackage{color}
\usepackage{multirow}
\usepackage{float}

\title[Warping protoplanetary discs]{Warping a protoplanetary disc with a planet on an inclined orbit}

\author[R. Nealon et al.]{\parbox{\textwidth}{
Rebecca Nealon$^{1}$\thanks{E-mail: rebecca.nealon@leicester.ac.uk},
Giovanni Dipierro$^{1}$,
Richard Alexander$^{1}$,
Rebecca G. Martin$^{2}$
\\
and Chris Nixon$^{1}$}\vspace{0.2cm}
\\
$^{1}$Department of Physics and Astronomy, University of Leicester, University Road, Leicester, LE1 7RH, UK\\
$^{2}$Department of Physics and Astronomy, University of Nevada, Las Vegas, 4505 South Maryland Parkway, Las Vegas, NV 89154, USA
}

\date{Accepted XXX. Received YYY; in original form ZZZ}

\pubyear{2017}

\begin{document}
\label{firstpage}
\pagerange{\pageref{firstpage}--\pageref{lastpage}}
 \maketitle

\begin{abstract}
Recent observations of several protoplanetary discs have found evidence of departures from flat, circular motion in the inner regions of the disc. One possible explanation for these observations is a disc warp, which could be induced by a planet on a misaligned orbit. We present three-dimensional numerical simulations of the tidal interaction between a protoplanetary disc and a misaligned planet. For low planet masses we show that our simulations accurately model the evolution of inclined planet orbit (up to moderate inclinations). For a planet massive enough to carve a gap, the disc is separated into two components and the gas interior and exterior to the planet orbit evolve separately, forming an inner and outer disc. Due to the inclination of the planet, a warp develops across the planet orbit such that there is a relative tilt and twist between these discs. We show that when other parameters are held constant, the relative inclination that develops between the inner and outer disc depends on the outer radius of the total disc modelled. For a given disc mass, our results suggest that the observational relevance of the warp depends more strongly on the mass of the planet rather than the inclination of the orbit.
\end{abstract}




\section{Introduction}
Planets form in cold discs of dust and gas orbiting young stars, and the gravitational interaction between forming planets and their parent disc plays a critical role in shaping young planetary systems \citep[e.g.][]{Kley:2012of,Baruteau:2014hb}. Planet-disc interactions may also create observable features, shown by high-resolution observations that have revealed a wealth of structures in many protoplanetary discs. These include spirals \citep[e.g.][]{Benisty:2015na,Stolker:2016ck,Stolker:2017of}, concentric rings \citep[e.g. HL Tau,][]{ALMA:2015ng} and large-scale asymmetries \citep[e.g.][]{vanderMarel:2018ve,Isella:2013ng,Perez:2014bq}. Such observations offer important new insights into the processes that shape the formation and early evolution of planetary systems.

Recent observations have also identified a number of protoplanetary discs with asymmetric emission that is strongly suggestive of warped or tilted disc structures. Near-IR observations show non-axisymmetric moving shadows in scattered light, as observed in TW Hya \citep{Debes:2017fk} and HD 135344B \citep{Stolker:2016ck}. In the case of TW Hya, this feature may be explained by a warped or tilted inner disc casting a shadow on the outer disc, such that the shadow moves with the precession of the warped inner disc \citep{Debes:2017fk,Poteet:2018be}. Molecular line profiles in several protoplanetary discs have also been shown to be inconsistent with flat, circular discs, and instead require significant vertical or radial gas motion (e.g., HD 100546, \citealt{Walsh:2017ic,Booth:2018ng}; RY Lup, \citealt{Arulanantham:2018og}), while (sub-)mm continuum observations suggest significant misalignments between the inner ($\lesssim1$~au) and outer ($\gtrsim 10$~au) disc in a number of different objects \citep[e.g.][]{vanderMarel:2015ne,vanderMarel:2018ve,Ansdell:2016gh}. In some cases the relative inclination between the inner and outer disc components may be quite large; $\sim$72$^{\circ}$ for HD 100453 \citep{Benisty:2017kq,Min:2017od}, $\sim$45$^{\circ}$ for AA Tau, $\sim$80$^{\circ}$ for HD 100546 \citep{Walsh:2017ic} and $\sim$30$^{\circ}$ for DoAr 44 \citep{Casassus:2018te}.

The deviation from planar, Keplerian motions observed in these discs is most readily explained by a gravitational perturbation from either a stellar or planetary companion. The asymmetric nature of the observed features suggests that the gravitational interaction is likely to be complicated by inclined or eccentric orbits \citep[e.g.][]{Ragusa:2017nv,Price:2018pf}. Stellar-mass binary companions are ruled out by current observations in most of these systems, but sub-stellar or planetary-mass perturbers on eccentric or inclined orbits may offer a self-consistent explanation for the observed disc structures \citep{Ruiz:2016ne}. Indeed, there is already circumstantial evidence for planetary companions in some of these systems. HD100546 has long been thought to host an embedded giant planet \citep[e.g.][]{Grady:2005vq,Acke:2006vs,Pinilla:2015ci}, perhaps on an inclined orbit \citep{Quillen:2006bv}. Recent observations of shocks in SO line emission support this hypothesis \citep{Booth:2018ng}. Similarly, in TW Hya the inner disc cavity is suggestive of dynamical clearing by a planetary companion \citep{Calvet:2002vr,Andrews:2016bw}, while recent observations also point towards the presence of one or more planets in the outer disc \citep{Tsukagoshi:2016hf,Uyama:2016ja}. Measurements of the disc inclination in TW Hya also hint at a small warp or misalignment, as observations of the inner disc ($\ll1$~au) typically infer inclinations that differ by a few degrees from those measured at large radii \citep[$\gtrsim 10$~au;][]{Qi:2004bw,Pontoppidan:2008ge,Hughes:2011gw}.

Although current planet formation theory assumes that protoplanetary discs are flat, a co-planar planet may become inclined or eccentric during or after formation. For example, planet-planet interactions can move a planet to an inclined orbit \citep{Nagasawa:2008pj}. Eccentricity growth can be driven through similar mechanisms \citep[e.g.][]{Papaloizou:2001ng} as well as planet-disc interactions \citep{Goldreich:2003er,Dangelo:2006ud,Teyssandier:2017uh}. Using an analytical prescription, \citet{Thommes:2003hb} have further demonstrated that exchanges between eccentricity and inclination between two planets in resonance can increase the inclination of both planets. As interactions energetic enough to drive strong misalignments are also violent enough to eject a planet, we note that modest inclinations are likely to be favoured by scattering events. If planet formation occurs early \citep[e.g.][]{Nixon:2018hf},  the disc may still be accreting material with potentially varying angular momentum directions \citep{Bate:2010nh}, and thus could be misaligned to the average disc plane. Additionally, protoplanetary discs are relatively thick (with aspect ratio $H/R \simeq $ 0.05--0.1, where $H$ is the disc scaleheight and $R$ the cylindrical radius) such that modest planet inclinations (up to several degrees) may naturally occur during the formation process.

\vspace{1cm}

Early theoretical work demonstrated that the evolution of the planet's orbit is determined by its interaction with the disc \citep{Goldreich:1979ir,Goldreich:1908nf}. Numerical simulations of planet-disc interactions have primarily focused on the planet's motion, with N-body calculations of embedded planets \citep[e.g.][]{Papaloizou:2000ks} paving the way for three-dimensional simulations. For planets slightly misaligned to the disc mid-plane, the motion of the planet is characterised by radial migration and inclination damping \citep{Tanaka:2002of,Tanaka:2004od}. \citet{Bitsch:2011ne} and \citet{Ayliffe:2010ow} demonstrated that the direction of migration is constrained by the choice of thermodynamics (i.e. whether or not radiative effects are included). \citet{Ayliffe:2010ow} showed that different numerical modelling of planetary accretion (e.g. as a sink particle, with softening or with a surface) led to different migration rates, where a more accurate migration rate was achieved with the smallest accretion radius as this permitted better modelling of the flow around the planet. \citet{Bitsch:2013pf} investigated the effect of viscosity on the migration rate, demonstrating that as low viscosity discs have low temperatures, they behaved in a similar way to isothermal discs, promoting inward migration. Independent of the direction of radial migration, the inclination of these planets was always damped \citep{Bitsch:2011ne}.

%
%
%

More recently, \citet{Bitsch:2013hg} and \citet{Xiang-Gruess:2013fg} studied the response of a disc driven by a misaligned planet. They found that planets massive enough to carve a gap can separate the disc into two components, both of which tilted away from the initial mid-plane of the disc toward the orbit of the planet. \citet{Xiang-Gruess:2013fg} considered a wide range of planet masses and inclinations, ranging from $1 \rm M_J$ to $6 \rm M_J$ (where $\rm M_J$ is the mass of Jupiter) and 10$^{\circ}$ to 80$^{\circ}$ respectively. They demonstrated that when the planet was massive enough to open a gap, the inclination took longer to damp. In their simulations the inner disc tilted faster than the outer disc, generating relative misalignments of up to 15$^{\circ}$ after 200 planet orbits. Similarly, isothermal simulations by \citet{Bitsch:2013hg} considered the evolution of the tilt between the inner and outer disc; after 400 planet orbits larger relative misalignments were observed for smaller initial planet inclinations. \citet{Bitsch:2013hg} interpreted that lower inclination planets have a stronger effect on the disc, leading to larger disc tilts for a fixed simulation time. Importantly, their simulations demonstrated that the disc can achieve an inclination greater than the planet. The evolution of higher mass planets at inclinations of up to 20$^{\circ}$ have also been studied by \citet{Marzari:2009of}, motivated by planets that may have undergone scattering events. In agreement with \citet{Xiang-Gruess:2013fg} their results suggest that strongly inclined planets can only carve a gap once their inclination has damped to a threshold determined by the planet mass \citep[see fig. 5 of][]{Xiang-Gruess:2013fg}. \citet{Cresswell:2007ss} further demonstrated that the eccentricity of an inclined planet is generally damped, in agreement with linear theory.

\citet{Arzamasskiy:2017ic} extended these studies by considering higher inclination planets and generating synthetic scattered light observations from a simulation of a disc with an inclined planet on a fixed orbit. At higher inclinations the planet spends a significant portion of its orbit outside of the disc, reducing planet-disc interactions to the instant where the planet passes through the disc. This process is usually analytically treated by a dynamical friction or aerodynamic drag approach \citep{Rein:2012so}. The synthetic observations generated by \citet{Arzamasskiy:2017ic} found little difference between the features formed by a co-planar or misaligned planet \citep[fig. 9,][]{Arzamasskiy:2017ic}. Common to all the numerical simulations that have been conducted to date is a restricted radial extent of $\lesssim 4$ times the semi-major axis of the planet (in many cases, this is equivalent to $\lesssim20$~au). However, protoplanetary discs are typically much more extended, with typical sizes of hundreds of au \citep{Andrews:2007pr,Ansdell:2018vi}. As we shall show, the choice of outer disc radius can have a significant impact on the response of the disc to the inclined planet and on the relative inclination that develops between the inner and outer disc.

In this paper we use three-dimensional numerical simulations to extend previous investigations of inclined planet-disc interactions, focusing on the disc structure that can be formed by a planet on a modestly inclined orbit (i.e. $i < 3\times H/R$, where $i$ is the inclination of the planet). While asymmetric disc features may be formed in discs around binary stars \citep[e.g.][]{lubow_ogilvie_2000,Price:2018pf} or planets in systems with misaligned binaries \citep{Lubow:2016nw}, our observational motivations constrain us to exclusively consider single star systems. We make use of the smoothed particle hydrodynamics code \textsc{Phantom} \citep{Phantom}. In Section~\ref{section:nm} we describe our numerical methods and the common features in our simulations. In Section~\ref{section:linear} we benchmark our numerical code against two tests outlined by \citet{Arzamasskiy:2017ic}, demonstrating that \textsc{Phantom} can accurately model the radial migration and inclination damping of a low mass planet. For higher mass planets, Section~\ref{section:warped} begins with a comparison to a previous simulation by \citet{Xiang-Gruess:2013fg}, subsequently demonstrating the importance of a large radial extent for such simulations. We then focus more broadly on the response of a disc to an inclined, massive planet by measuring the averaged tilt and twist of disc segments carved by the planet. In Section~\ref{section:disc} we interpret these results by considering the relevant time-scales that describe the behaviour of the disc, and Section~\ref{section:concs} summarises our findings.

\begin{table*}
	\caption{Summary of parameters used for our simulations. Multiple values indicate multiple simulations. $N$ is the number of particles, $i$ the initial inclination of the planet with respect to the mid-plane of the disc (in degrees), $m$ is the mass of the planet (in Jupiter masses), $L_{\rm p}/L_{\rm out}$ is the ratio of the angular momentum of the planet to the angular momentum of the gas exterior to the planet orbit and $R_{\rm out}$ the outer disc radius (in~au). The final time of the simulation, $t_f$, is measured in orbits of the planet at 5 au, $q$ determines the power-law slope for the sound speed (Equation~\ref{equation:sound_speed}), $R_{\rm acc}$ is the accretion radius of the planet where the Hill radius is defined in Equation~\ref{equation:hill_radius}. $\langle h \rangle/H$ is the shell averaged smoothing length to scaleheight ratio measured at the location of the planet, an indication of the resolution in each simulation (with a lower value indicating higher resolution).}
	\begin{tabular}{c|c|c|c|c|c|c|c|c|c}
		\hline
		Name & $N$ & $i$ & $m$ & ${L}_{\rm p}/{L}_{\rm out}$ & $R_{\rm out}$ & $t_f$ & $q$ & $R_{\rm acc}$  & $\langle h \rangle/H$ \\
		\hline
		G Low & 1.25$\times10^5$ & 1.62-19.42 & 0.40 & 0.18  & 20 & 500 & 0 & 0.25 $R_{\rm Hill}$ & 0.50\\
		G Med & 1.0$\times 10^6$ & 1.62-19.42 & 0.40 & 0.18 & 20 & 100 & 0 & 0.25 $R_{\rm Hill}$ & 0.23\\
		G High & 8.0$\times 10^6$ & 1.62-19.42 & 0.40 & 0.18 & 20 & 20 & 0 & 0.25 $R_{\rm Hill}$ & 0.12\\
		G2 & 1.0$\times 10^6$ & 1.62 & 0.40 & 0.18 & 20 & 100 & 0 & 0.125 $R_{\rm Hill}$ & 0.23\\
		G3 & 1.0$\times 10^6$ & 1.62 & 0.40 & 0.18 & 20 & 100 & 0 & 0.40 $R_{\rm Hill}$ & 0.23\\
		\hline
		L1 Low & 1.25$\times 10^5$ & 20.0 & 6.5 & 2.69 & 20 & 200 & 0.25 & 0.1 au & 0.60\\
		L1 Med & 1.0$\times 10^6$ & 20.0 & 6.5 & 2.69 & 20 & 200 & 0.25 & 0.1 au & 0.23\\
		L2 & 3.0$\times 10^5$ & 20.0 & 6.5 & 0.60 & 50 & 200 & 0.25 & 0.1 au & 0.62\\
		L3 & 7.3$\times 10^5$ & 20.0 & 6.5 & 0.20 & 100 & 200 & 0.25 & 0.1 au & 0.60\\
		L Low & 4.3$\times 10^5$ & 2.15, 4.30, 12.89 & 0.13, 1.3, 6.5 & 0.01 - 0.63 & 50 & 500 & 0.25 & 0.1 au & 0.50\\
		L Med & 3.4$\times 10^6$ & 2.15, 4.30, 12.89 & 0.13, 1.3, 6.5 & 0.01 - 0.63 &  50 & 200 & 0.25 & 0.1 au & 0.28\\
		\hline
	\end{tabular}
	\label{tab:sims_summary}
\end{table*}

\section{Numerical Method}
\label{section:nm}
We use the smoothed particle hydrodynamics (SPH) code \textsc{Phantom} \citep{Phantom}. \textsc{Phantom} has been used extensively to investigate the evolution of accretion discs, including modelling binary driven effects \citep[e.g.][]{nixon_2013,Martin:2014aa,Martin:2014bb,Dogan:2015rt,Price:2018pf} and planets orbiting a binary system \citep{Martin:2014bb,Martin:2016qf}. The evolution of the disc in response to an embedded planet has also been considered \citep{Dipierro:2015od,Dipierro:2016ss,Ragusa:2017nv}. Additionally, the interaction between a disc and a misaligned planet has already been studied using SPH by \citet{Xiang-Gruess:2013fg} and \citet{Marzari:2009of}.

The tidal interaction of an inclined planet with its parent protoplanetary disc leads to the excitation of disturbances or warps that propagate via bending waves \citep{pap_pringle_1983,Papaloizou:1995pn}. This regime of warp propagation occurs whenever the $\alpha$ parameter \citep{shakura_sunyaev} is smaller that the disc aspect ratio, i.e. $\alpha < H/R$, where $H$ is the disc scaleheight and $R$ is the cylindrical radius. This condition is readily satisfied in typical protoplanetary discs \citep{Flaherty:2015is,Flaherty:2017do,Pinte:2016oa}. Theoretical studies in the linear regime have shown that these waves  propagate across the disc at half of the local sound speed \citep[e.g.][]{Papaloizou:1995pn} and are then damped non-locally through the action of viscosity. Numerical studies with three-dimensional simulations have confirmed these results, including SPH simulations using \textsc{Phantom} \citep{lodato_2010,Nealon:2015fk,Nealon:2016lr}. A test confirming accurate warp propagation in the wave-like regime with \textsc{Phantom} is presented in fig. 2 of \citet{Nealon:2015fk}.

In our simulations the planet is started at an initial distance from the star of $a=$ 5~au and the central star has a mass M$_*$~=~1M$_{\odot}$. The total disc mass between $R_{\rm in} = 0.1$~au and $R_{\rm out} = 100$~au is 0.01M$_{\odot}$. While our initial conditions prescribe circular orbits, these are not enforced during the simulation as the planet is free to move. Time-scales are quoted in planet orbits at the planet's initial location and our remaining parameters are summarised in Table~\ref{tab:sims_summary}. We perform our simulations at multiple resolutions to check for convergence and our results are presented with the highest resolution in each section. 

We adopt an isothermal equation of state, with the sound speed in the disc described by
\begin{align}
c_{\rm s}=c_{\rm s,0} \left(\frac{R}{R_0}\right)^{-q}, \label{equation:sound_speed}
\end{align}
and the aspect ratio is given by
\begin{align}
\frac{H}{R}= 0.05 \left(\frac{R}{R_0}\right)^{1/2-q}.
\end{align}
In the above equations $q$ determines the radial profile of the sound speed and we use $R_0 = 1$~au. The parameter $c_{\rm s,0}$ is determined from hydrostatic equilibrium (i.e. $H/R = c_{\rm s}/v_{\rm K}$, where $v_{\rm K}$ is the Keplerian velocity) at $R=R_0$. The surface density profile $\Sigma(R)$ is given by \citep{Pringle:1981fo}
\begin{align}
\Sigma(R)=\Sigma_0 \left(\frac{R}{R_0} \right)^{-p} \left(1-\sqrt{\frac{R_{\rm in}}{R}}\right), \label{equation:sigma}
\end{align}
and the scaling $\Sigma_0$ is determined by the disc mass. We adopt $p=1$ for all our simulations, but note that due to the taper used at the inner edge of the disc the actual power-law slope at the location of the planet is measured from the simulation to be $\approx 0.95$.

The planet and star in our simulations are modelled using sink particles \citep[see Section 2.8 of][]{Phantom}. Sink particles are allowed both to accrete and migrate as a consequence of their interactions with the disc and their mutual gravitation interaction \citep{Bate:1995fi}. The accretion radius of the central star is set at 0.1~au in all our simulations, while we vary the accretion radius of the planet (see Table~\ref{tab:sims_summary}). No gravitational softening of sink particles is employed for these simulations.

Viscosity is modelled using the \citet{shakura_sunyaev} $\alpha$ parameter, with $\alpha=10^{-3}$ imposed for all of our simulations. As the planet in our simulations spends time above and below the disc (and sometimes in a gap) we use a resolution independent viscosity implementation. Here the viscous terms are calculated from the Navier-Stokes equations \citep{Flebbe:1994lr} and the kinematic viscosity is set by the disc scaleheight $H$ and sound speed $c_{\rm s}$ such that
\begin{align}
\nu = \alpha c_{\rm s} H.
\end{align}
This calculation is made for each particle, using the disc properties and chosen $\alpha$. This method is documented in Section 2.7.2 of \citet{Phantom} and has been used previously with \textsc{Phantom} \citep[e.g.][]{facchini_2013}. Artificial numerical viscosity is included with a switch to accurately capture shocks, with $\alpha_{\rm AV} \epsilon [0.01,1.0]$ and $\beta_{\rm AV}=2.0$ \citep[see][]{Phantom}.

In Section~\ref{section:linear} we consider the motion of a planet with a low planet mass. In order to compare to existing analytical solutions, we use a globally isothermal equation of state by setting $q=0$. From Section~\ref{section:warped} onwards we use a locally isothermal equation of state with $q=0.25$, consistent with observations \citep{Andrews:2007pr,Williams:2011he}.

\section{Linear Regime}
\label{section:linear}
In this section present a numerical test of the motion of a low mass, inclined planet in a thin disc. We follow the procedure outlined by \citet{Bitsch:2013hg} and \citet{Arzamasskiy:2017ic}, focusing on the radial migration and inclination damping rates. To ensure that the gravitational perturbations from the planet are weak we set the mass to be one tenth the thermal mass. Assuming zero eccentricity, the thermal mass is defined as
\begin{align}
\frac{m_{\rm thermal}}{M_*}=3 \left( \frac{H}{R}\right)^3,
\end{align}
with $H/R$ evaluated at the location of the planet. The thermal mass corresponds to the mass where the disc scaleheight $H$ is equal to the Hill radius of the planet, defined as
\begin{align}
R_{\rm Hill} = a\left(\frac{m}{3M_*}\right)^{1/3}, \label{equation:hill_radius}
\end{align}
where $a$ is the semi-major axis. For the globally isothermal simulations we consider here, $H/R = 0.113$ so $m_{\rm thermal}=0.4 \rm M_{J}$. In this section we set $R_{\rm out} = 20$~au and consider eight initial planet inclinations between $i=1.62^{\circ} - 19.42^{\circ}$, equivalent to 0.25$\times H/R - 3\times H/R$.

\begin{figure*}
	\includegraphics[width=\textwidth]{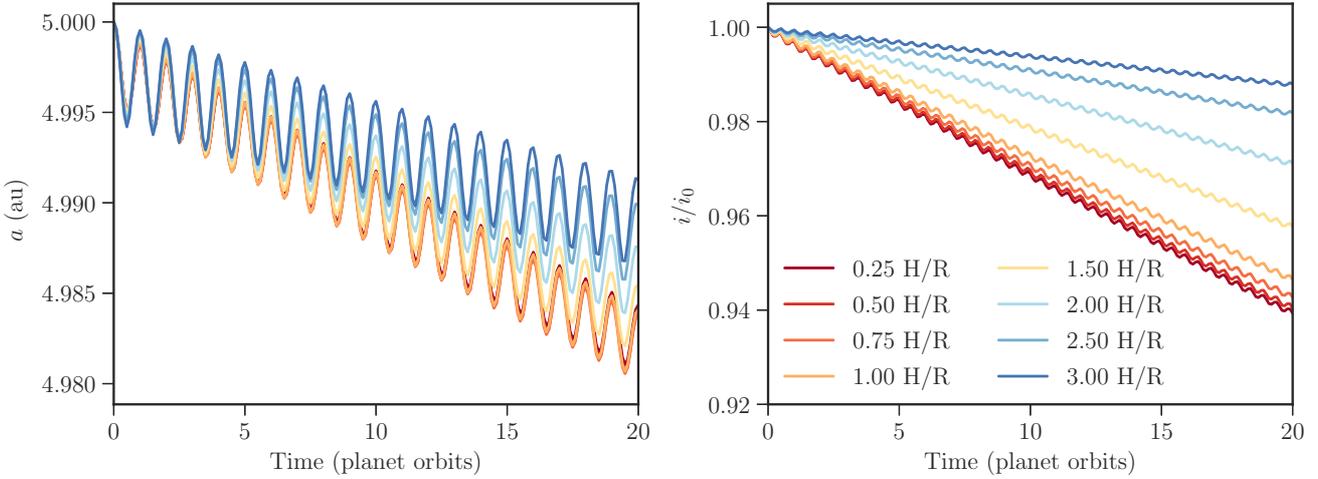}
    \caption{Evolution of the semi-major axis $a$, and the orbital inclination $i$, (relative to the initial inclination $i_0$) for a low mass planet on an inclined orbit (G High simulations). When calculating the averaged radial migration and inclination damping rates we use data from $t$ =10-20 planet orbits only to avoid initial transients. Our lower resolution calculations suggest that the behaviour shown continues for $\sim$100 orbits. The small magnitude oscillations that occur on the orbital time-scale are caused by the orbital variations of the torque on the planet. \label{fig:movement}}
\end{figure*}

Fig.~\ref{fig:movement} shows the evolution of the semi-major axis and the inclination of the planet for these simulations (labelled G High). The motion of the planet can be expressed by its acceleration due to interactions with the disc. \citet{Burns:1976of} decompose the acceleration of the planet into orthogonal components $R$, $T$ and $N$ such that
\begin{align}
\ddot{\mathbf{r}}=R\mathbf{e_R} + T\mathbf{e_T} + N\mathbf{e_N}. \label{equation:acceleration_decomposition}
\end{align}
These are the components due to the planet's interaction with the disc (i.e. these do not include contributions from the star). Here $\mathbf{r}$ is the position vector, $\mathbf{e_R}$ the unit vector in the radially outward direction, $\mathbf{e_T}$ in the direction of the planet's velocity (positive in the direction of motion of the planet) and $\mathbf{e_N}$ normal to the plane of the orbit (i.e. in the direction of $\mathbf{e_R} \times \mathbf{e_T}$). The radial migration of a planet is expressed by \citet{Burns:1976of}
\begin{align}
\frac{da}{dt}=2 \frac{a^{3/2}}{\sqrt{GM_*(1-e^2)}} \left[e R \sin{f} + T(1 + e\sin{f}) \right],
\label{equation:radial_migration_burns}
\end{align}
where $G$ is the universal gravitational constant, $e$ the eccentricity of the planet, $f$ is the true anomaly and $R$, $T$ are the directional accelerations defined in Equation~\ref{equation:acceleration_decomposition}. In this form it is clear that only the acceleration components in the plane of the orbit ($R$ and $T$) are able to change the semi-major axis of the orbit. Similarly, the inclination damping driven only by interactions between the disc and planet is \citep{Burns:1976of}
\begin{align}
\frac{di}{dt}=\frac{r N \cos{\theta}}{{L}_{\rm P}},
\label{equation:inclination_damping_burns}
\end{align}
where $\theta$ denotes the position angle between the line of nodes and the planet along its orbit around the star and ${L}_{\rm P}$ is the magnitude of the planet specific angular momentum, defined by ${L}_{\rm P}=\sqrt{GM_* a (1-e^2)}$. Here, only the component of the acceleration perpendicular to the plane of the orbit ($N$) is able to damp the inclination.

We consider in turn the radial migration in Section~\ref{subsection:radial_migration} and inclination damping in Section~\ref{subsection:inclination_damping}. As the planet's orbital evolution is included in our simulations, we are able to evaluate Equations~\ref{equation:radial_migration_burns} and \ref{equation:inclination_damping_burns} using two methods. First, we use the accelerations experienced by the planet from the disc-planet interactions (i.e. the right hand side of Equations~\ref{equation:radial_migration_burns} and \ref{equation:inclination_damping_burns}). Second, we use the evolution of the planet's location to calculate the radial migration and inclination damping rates directly (i.e. the left hand side of Equations~\ref{equation:radial_migration_burns} and \ref{equation:inclination_damping_burns}).

In both cases, to avoid any transients due to our initial conditions we average these results from 10 to 20 orbits of the planet. As these simulations use adaptive timestepping, the measurements of the acceleration are at unequally spaced time intervals. We thus use the following formula to calculate the weighted average over multiple orbits, where for the example quantity $Y$
\begin{align}
\langle Y \rangle=\frac{\sum Y_i \Delta t_i}{\sum \Delta t_i},
\end{align}
where $\Delta t_i$ is the interval between each measurement and $\langle Y\rangle$ is the time averaged value. We evaluate the migration and inclination damping rates in terms of their respective time-scales \citep{Tanaka:2004od},
\begin{align}
t_{\rm mig} &= \Omega_k^{-1} \left( \frac{H}{R} \right)^2 \left( \frac{m}{M_*} \right)^{-1} \left( \frac{\Sigma R^2}{M_*}\right)^{-1},\\
t_{\rm inc} &= \left(\frac{H}{R} \right)^2 t_{\rm mig}.
\end{align}
Here $\Omega_k$ is the angular velocity of the planet and all properties are calculated at the location of the planet. In all the simulations shown here, the eccentricity of the planet remains below $\approx 1.0 \times 10^{-3}$ for the duration of the simulation.

\begin{figure*}
	\includegraphics[width=\textwidth]{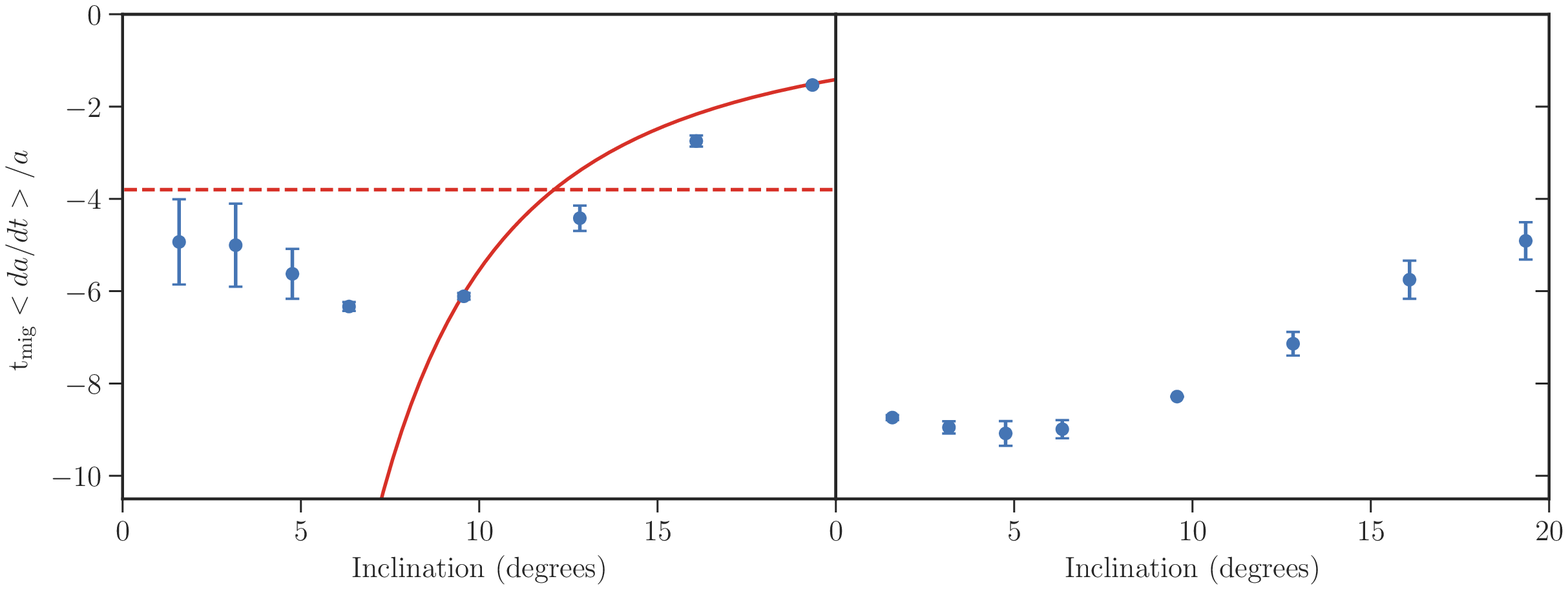}
    \caption{Radial migration rate for a planet in the linear mass regime at a range of different initial inclinations (G High). Left: The points are calculated from the accelerations experienced by the planet (right hand side of Equation~\ref{equation:radial_migration_burns}). The red lines indicate the inviscid analytical predictions, with the horizontal line described by Equation~\ref{equation:tanaka_2002} and the curved line by Equation~\ref{equation:rein_2012_rad} ($\Lambda$ =3.28). The trend with increasing inclination is consistent with \citet{Arzamasskiy:2017ic}. Right: The points are measured using the derivative of the motion of the planet (left hand side of Equation~\ref{equation:radial_migration_burns}). In both panels the uncertainties are derived from the difference between the G High and G med simulations. \label{fig:radial_migration}}
	\includegraphics[width=\textwidth]{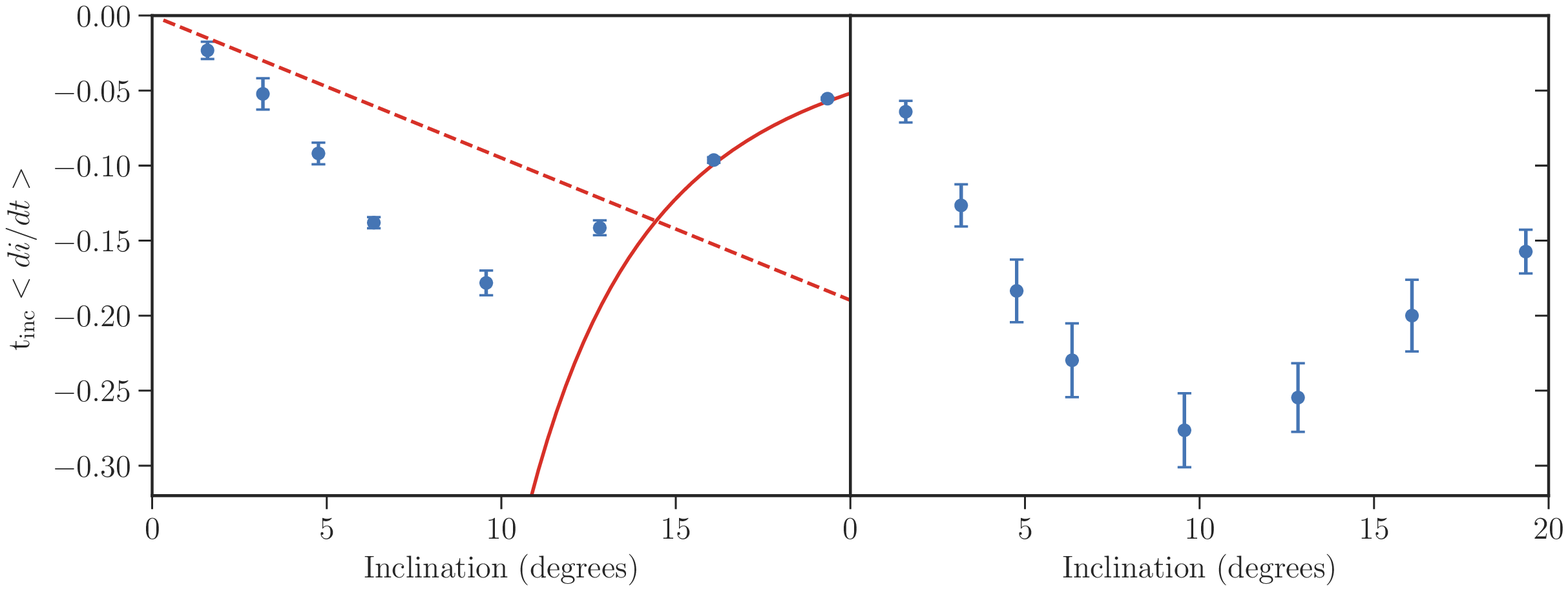}
    \caption{Inclination damping rate for a planet in the linear mass regime at a range of different initial inclinations (G High). Left: The points are calculated from the accelerations experienced by the planet (right hand side of Equation~\ref{equation:inclination_damping_burns}). The red lines indicate the inviscid analytical predictions, with the straight line described by Equation~\ref{equation:tanaka_2004} and the curved by Equation~\ref{equation:rein_2012_inc} ($\Lambda$ =1.66). The trend with increasing inclination is again consistent with \citet{Arzamasskiy:2017ic}. Right: The points are measured using the derivative of the motion of the planet (left hand side of Equation~\ref{equation:inclination_damping_burns}). In both panels the uncertainties are derived from the difference between the G High and G med simulations.\label{fig:inclination_damping}}
\end{figure*}

\subsection{Radial migration}
\label{subsection:radial_migration}
The left panel of Fig.~\ref{fig:radial_migration} shows scaled radial migration time-scale, calculated from the right hand side of Equation~\ref{equation:radial_migration_burns}. The uncertainties for each point are calculated from the difference between the rate measured in the G High and G Med simulations for each planet inclination. For small inclinations, we compare to the analytical form for the radial migration of a low mass planet that is aligned with the mid-plane, assuming an inviscid and globally isothermal disc \citep{Tanaka:2002of}:
\begin{align}
\frac{ \langle da/dt \rangle}{a} =-(2.7 + 1.1p)\, t_{\rm mig}^{-1}.
\label{equation:tanaka_2002}
\end{align}
Here $p$ is the power-law index for the surface density profile in Equation~\ref{equation:sigma}, measured at the location of the planet. The right hand side of Equation~\ref{equation:tanaka_2002} is included in Fig.~\ref{fig:radial_migration} with a horizontal dashed red line. At small inclinations ($i < H/R$) the migration rate is almost constant, in agreement with the prediction from Equation~\ref{equation:tanaka_2002}, albeit slightly faster than predicted.

At larger inclinations ($i \gtrsim H/R$) the planet-disc interaction only occurs as the planet briefly passes through the disc twice per orbit. The planet dynamics in this regime can be described by the dynamical friction experienced by the planet once it crosses the disc plane along its orbit \citep{Rein:2012so,Teyssandier:2013og}. For these inclinations, we compare our results to the prediction from \citet{Rein:2012so},
\begin{align}
\frac{\langle da/dt \rangle}{a}=- \frac{2 \Lambda}{\sin(i/2) \sin(i)} \left( \frac{H}{R} \right)^2 t_{\rm mig}^{-1}.
\label{equation:rein_2012_rad}
\end{align}
Here $\Lambda$ behaves like a Coulomb logarithm and is expected to be between 1$\sim$10 \citep{Rein:2012so}. As in \citet{Arzamasskiy:2017ic} we treat this as a fitting parameter, using our simulation at $i=19.42^{\circ}$ to constrain $\Lambda=3.28$ (the curved red line, Fig.~\ref{fig:radial_migration}). The trend with increasing inclination of Equation~\ref{equation:rein_2012_rad} is broadly replicated for larger inclinations. Our fitting parameter $\Lambda$ is within 25\% of the value used in \citet{Arzamasskiy:2017ic}, and the profile of migration rate with increasing initial inclination is consistent with fig. 4 of \citet{Arzamasskiy:2017ic} and the isothermal case shown in fig. 3 of \citet{Bitsch:2011ne}. As the inclination increases from linear to non-linear, we find the fastest migration rate occurs at $i\approx H/R$, similar to \citet{Arzamasskiy:2017ic}.

Fig.~\ref{fig:radial_migration}, right panel shows the radial migration calculated using the actual change in position of the planet. We find the same trend as a function of inclination as in the left panel, but with a constant offset such that the planet is moving faster than predicted by the previous method. This discrepancy is due to our modelling of accretion on to the planet --- by simply accreting particles that are within $R_{\rm acc}$ we do not model the flow near to the planet as accurately as possible. A more realistic scenario sees some particles entering the Hill sphere, interacting with the planet and then exiting \citep{Ayliffe:2010ow}. As this gas-planet interaction close to the planet leads to angular momentum exchange, simply accreting those particles will adversely affect the migration rate (see Appendix~\ref{section:racc_study}).

\subsection{Inclination damping}
\label{subsection:inclination_damping}
Fig.~\ref{fig:inclination_damping}, left panel shows the scaled inclination damping time-scale, calculated from the right hand side of Equation~\ref{equation:inclination_damping_burns}. We follow the same process as for the radial migration, comparing to analytical predictions for small and large inclinations. For small inclinations, we use the analytical model of \citet{Tanaka:2004od}:
\begin{align}
\langle di/dt \rangle= -0.544 \frac{i}{ t_{\rm inc}}.
\label{equation:tanaka_2004}
\end{align}
Equation~\ref{equation:tanaka_2004} is shown as a straight line in Fig.~\ref{fig:inclination_damping}. At these low inclinations the measurements from the accelerations on the planet again follow the broad trend with inclination, although slightly overestimating the predicted damping rate.

At larger inclinations we refer to \citet{Rein:2012so}, with the inclination damping
\begin{align}
\langle di/dt \rangle= -\frac{\Lambda}{\sin^3(i/2)} \left( \frac{H}{R} \right)^4 t_{\rm inc}^{-1}.
\label{equation:rein_2012_inc}
\end{align}
Following \citet{Arzamasskiy:2017ic}, we fit the solid red line in Fig.~\ref{fig:inclination_damping} to our simulation at $i=19.42^{\circ}$ with $\Lambda~=~1.66$ (within 12\% of that found by \citealt{Arzamasskiy:2017ic}). The value of $\Lambda$ for the inclination damping time-scale is smaller with respect to that used for the radial migration rate, consistent with \citet{Arzamasskiy:2017ic}. The distinction in $\Lambda$ values is attributed to the anisotropy of the disc \citep{Arzamasskiy:2017ic,Rein:2012so}. Our profile of inclination damping as a function of initial planet inclination is consistent with fig. 5 of \citet{Arzamasskiy:2017ic} and the general profiles in fig. 1 of \citet{Bitsch:2011ne} and fig. 5 of \citet{Bitsch:2013hg}. Additionally, at high inclinations we identify the same trend of $\langle di/dt \rangle \propto i^{-3}$ as \citet{Arzamasskiy:2017ic}, contrary to \citet{Bitsch:2011ne} and \citet{Cresswell:2007ss}. \citet{Arzamasskiy:2017ic} suggest this discrepancy is due to \citet{Bitsch:2011ne} only using smaller inclinations than considered in their work.

Fig.~\ref{fig:inclination_damping}, right panel shows the same inclination damping as measured using the motion of the planet as a function of time. As with the radial migration time-scale, due to our choice of accretion radius the inclination damping profile with increasing inclination is replicated with an offset (see Appendix~\ref{section:racc_study}).

The comparison between our simulations and the analytically-derived radial migration and inclination damping rates demonstrate that \textsc{Phantom} is able to model the orbit of an inclined planet accurately in the the linear planet mass regime. When we measure the radial migration and inclination damping rates from the accelerations of the planet (left hand panels), our results are consistent with previous results using grid codes by \citet{Bitsch:2011ne} and \citet{Arzamasskiy:2017ic}. When we instead measure these rates from the actual motion of the planet (right hand panels) we find slightly faster radial migration and inclination damping time-scales, but this due to our choice of accretion radius (Appendix~\ref{section:racc_study}). We now extend our simulations to larger mass planets that may be capable of warping the disc.

\section{Warping protoplanetary discs}
\label{section:warped}
In this section we investigate how much the disc structure may be altered by a misaligned planet. Here we adopt a locally isothermal equation of state (the `L' prefix in Table~\ref{tab:sims_summary}), such that $c_{\rm s} \propto R^{-1/4}$ and $H/R \propto {R^{1/4}}$. These choices are motivated by observations of protoplanetary discs, where the temperature is typically described by $T \propto R^{-1/2}$  \citep{Kenyon:1987bg,Andrews:2007pr,Williams:2011he}. These simulations are also run for ten times longer than those in the previous section, so we utilise a larger accretion radius for the planet of 0.1~au ($R_{\rm acc} = 0.1\text{~au} < 0.4 R_{\rm Hill}$ in practise). As demonstrated in Fig.~\ref{fig:racc_convergence}, this slightly decreases the accuracy of the radial migration and inclination damping rates. While a larger accretion radius leads to an overestimate of the inclination damping rate (see Appendix~\ref{section:racc_study}), this does not affect the response of the disc significantly.

As we are concerned with the observational consequences of the planet on the disc, we focus on the evolution of the disc angular momentum vector characterised by the tilt $\beta(R,t)$ and twist $\gamma(R,t)$. Measuring these angles relative to the $z$ axis, the unit vector of the disc is expressed in spherical polar coordinates as \citep[e.g.][]{pringle_1996,nixon_2012}
\begin{align}
\boldsymbol{\ell}(R,t)=(\cos \gamma \sin \beta, \sin \gamma \sin \beta, \cos \beta).
\end{align}
In our analysis we set the total angular momentum vector $\mathbf{L}_{\rm total} = \mathbf{L}_{\rm disc} + \mathbf{L}_{\rm planet} + \mathbf{L}_{\rm star}$ to be coincident with the $z$ axis. The unit vector of the disc $\boldsymbol{\ell} (R,t)$ is evaluated by discretising the disc into radial annuli and azimuthally averaging the properties of the particles in each annulus \citep[as described in][]{lodato_2010}. The evolution in tilt and twist of the disc are used to infer the motion of the disc; a decrease in tilt indicates the disc is tilting towards the total angular momentum vector while a change in twist indicates the disc is precessing. The tilt and twist are simply estimated from $\boldsymbol{\ell} (R,t)$ by
\begin{align}
\beta(R,t)=\cos^{-1} \left(\boldsymbol{\ell}_z(R,t)\right),
\end{align}
\begin{align}
\gamma(R,t)=\tan^{-1} \left( \frac{\boldsymbol{\ell}_y(R,t)}{\boldsymbol{\ell}_x(R,t)} \right).
\end{align}
The disc is considered to be warped when the tilt or twist vary as a function of radius. The precession rate of the disc is inferred from the twist evolution as
\begin{align}
t_{\rm prec, \gamma}=2\pi \left( \frac{d \gamma}{dt} \right)^{-1}. \label{equation:tp}
\end{align}
The above approach assumes azimuthal symmetry around $\mathbf{L}_{\rm total}$, so we confirm that asymmetric features are not washed out by checking the three-dimensional density rendering of each simulation. To avoid any features generated by the tidal streams near the planet, we also do not consider any gas inside the gap created by the planet.

\begin{figure*}
	\includegraphics[width=\textwidth]{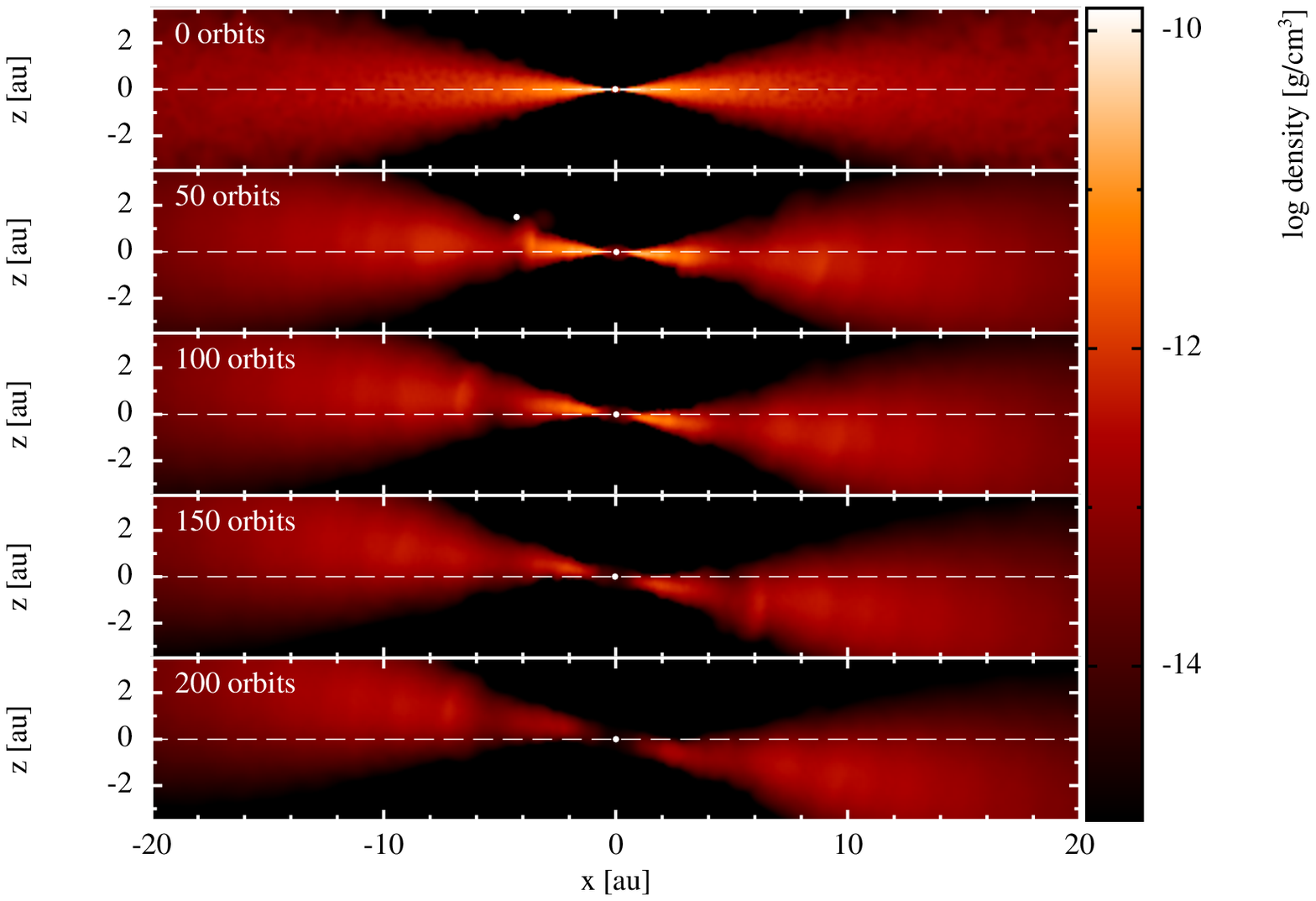}
    \caption{Density cross-section at $y$ = 0 of the simulation L1 Med, where $m=6.5 \rm M_J$, $R_{\rm out} = 20$~au and the planet and star are marked with white circles. The inclined planet carves a gap, separating the disc into an inner and outer disc. Both discs tilt up towards the planet orbit, with the inner disc moving more rapidly. The planet is only visible in these cross-sections when it happens to pass through the $y$ = 0 plane. \label{fig:pretty}}
\end{figure*}

\begin{figure*}
	\includegraphics[width=\textwidth]{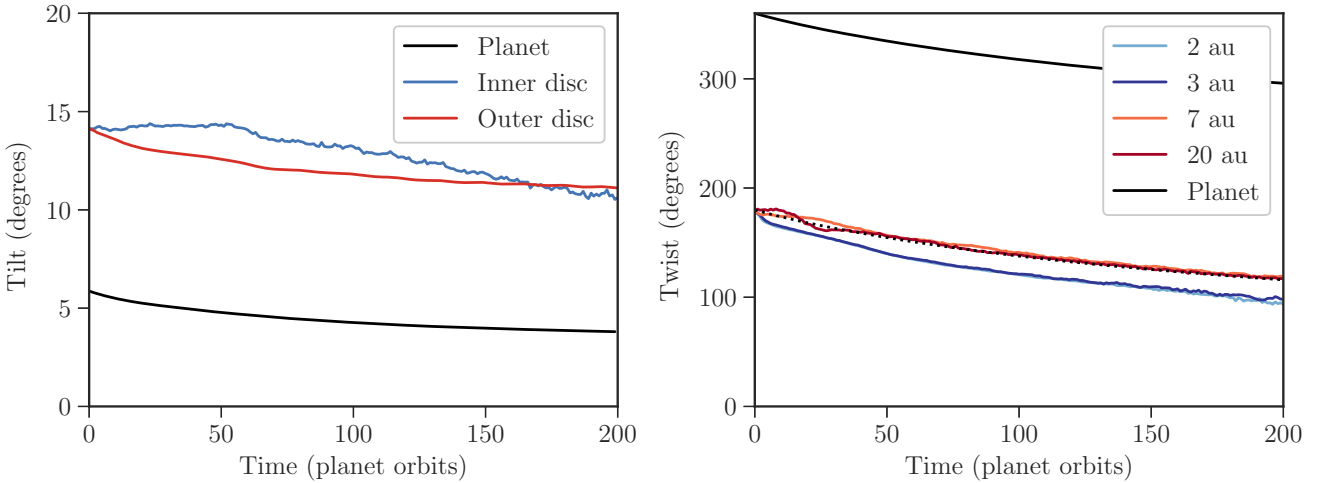}
    \caption{Evolution of the tilt and twist in the simulation in Fig.~\ref{fig:pretty}, measured with respect to $\mathbf{L}_{\rm total}$. The inner disc is shown in blues, the outer disc in reds and the planet shown in black. The initial disc-planet misalignment ($20^{\circ}$) is the sum of the planet and disc tilt. The difference in tilt and twist measured between the inner and outer disc in conjunction with the same rate of twist suggests that the disc is warped across the inner and outer disc but globally precesses as a solid body. As expected, the disc twists in a retrograde sense to the planet orbit. Due to accretion on to the star, the inner disc is less well resolved than the outer disc leading to a noisier measurement.\label{fig:matching_XGPap2013}}
\end{figure*}

\begin{figure*}
	\includegraphics[width=\textwidth]{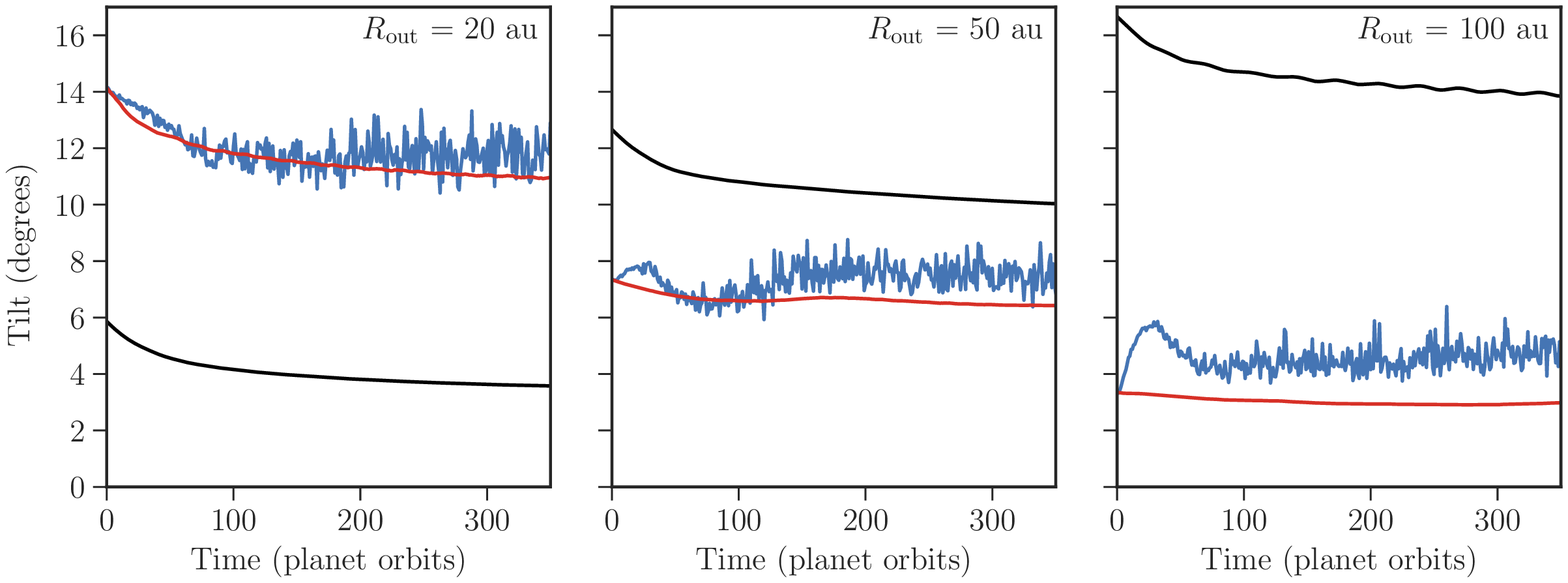}
    \caption{Increasing $R_{\rm out}$ alters the relative inclination measured between the gas interior and exterior to the planet orbit. The gas interior to the planet orbit is shown in blue, exterior in red and the planet inclination shown with black. In the left panel ($R_{\rm out} = 20$~au) the planet  has a lower inclination than the disc components, so $\mathbf{L}_{\rm p}$ is closer to $\mathbf{L}_{\rm total}$ and the planet has a greater angular momentum than the disc. As the outer radius is increased in the middle (50 au) and right (100 au) panels, the fraction of the total angular momentum in the disc increases such that $\mathbf{L}_{\rm total}$ is closer to $\mathbf{L}_{\rm disc}$ and further from $\mathbf{L}_{\rm p}$. This is demonstrated by the disc components having a smaller tilt than the planet when $R_{\rm out}$ is larger. With the largest outer radius, the total disc has much more angular momentum than the planet and so the outer disc does not respond appreciably to the influence of the inclined planet. The inner disc still does, leading to a greater relative inclination.\label{fig:outer_edge_resolution_test}}
\end{figure*}

As the response of a disc to the influence of an inclined planet has already been studied in depth \citep{Xiang-Gruess:2013fg,Bitsch:2013hg}, we begin with a comparison to one of the simulations by \citet{Xiang-Gruess:2013fg}. In Section~\ref{subsection:matching_xgpap} we demonstrate that this reference simulation is consistent with the previous results. We refer to the section of the disc interior to the planet orbit as the inner disc and the disc exterior as the outer disc. In Section~\ref{subsection:outeredge} we consider the effect of the choice of $R_{\rm out}$ on the evolution of the disc. We use the results from this section to inform the initial conditions for the simulations in Section~\ref{subsection:warping}, and in Section~\ref{subsection:warp_strength} we consider the combination of planet masses and inclinations that can drive significant warps.

\subsection{Reference case}
\label{subsection:matching_xgpap}
We begin by repeating the highest inclination of the `G Med' simulations, but altering the initial planet inclination to $i=20^{\circ}$ (corresponding to $\sim 5 \times H/R$, where $H/R = 0.075$ at the planet location). We increase the planet mass to 6.5 Jupiter masses, equivalent to five times $m_{\rm thermal}$ at 5~au in a locally isothermal disc. This simulation is run for 200 planet orbits with $N=1.25\times 10^5$ and $1.0\times10^6$ particles (L1 Low and High). The lower resolution of these two simulations has similar parameters to one of the simulations presented by \citet{Xiang-Gruess:2013fg}, where they used a slightly higher resolution of $2.0\times10^5$ particles (but find similar evolution in their resolution study, which includes a simulation with $1.0\times10^5$ particles). While \citet{Xiang-Gruess:2013fg} also use SPH for their simulation, they use an adapted locally isothermal equation of state to our simulation, altering accretion on to the planet.

Fig.~\ref{fig:pretty} displays the evolution of the density cross-section of simulation L1 Med from $t=0$ to $t=200$ planet orbits. Qualitatively, we observe similar behaviour to that found by \citet{Xiang-Gruess:2013fg}; as the planet inclination damps it carves the disc into an inner and outer region. Both regions tilt up towards the planet's orbital plane, with the inner disc initially tilting faster. As the inner and outer disc tilt away from the initial configuration, a relative inclination develops between them. This evolution of the disc is also demonstrated using a grid code by \citet{Bitsch:2013hg} at smaller inclinations (their fig. 3).

Fig.~\ref{fig:matching_XGPap2013} displays the tilt (left panel) and twist (right panel) relative to the total angular momentum vector over the first 200 orbits for the gas interior to the planet orbit (in blue), exterior to the planet orbit (in red) and the planet orbit (in black). We note that the planet is initally inclined at $20^{\circ}$ relative to the disc ($\beta_{\rm planet} + \beta_{\rm disc} = 20^{\circ}$), corresponding to a planet inclination of $\beta_{\rm planet} \sim 6^{\circ}$ and disc inclination of $\beta_{\rm planet} \sim 14^{\circ}$ from the total angular momentum vector (as shown in Fig.~\ref{fig:matching_XGPap2013}). The blue or red line is calculated by averaging the angular momentum vector of each annulus that contains particles inside or outside the planet orbit, respectively. As we shall demonstrate, because neither disc is itself warped this represents the behaviour of each disc near the planet orbit. Transients are washed out within the first $\sim$ten orbits. The decrease in tilt demonstrates that the disc and the planet momentum vectors tilt towards the total angular momentum vector of the system, while a small misalignment develops between the inner and outer disc ($\sim$0.5$^{\circ}$ by 200 orbits). In order to demonstrate solid body rather than differential precession, the twist (right panel) is measured at four locations in the disc; 2~au (light red), 3~au (dark red), 7~au (light blue) and 10~au (dark blue). The similarity of the twist and the evolution of the twist measured at the two inner radii and, independently, the two outer radii confirm that each section of the disc is precessing as a solid body (i.e. with the same $t_{\rm prec, \gamma}$, see Equation~\ref{equation:tp}). Due to its large mass, the planet (black line) precesses with the outer disc but is out of phase by 180$^{\circ}$ (the dotted black line). As $\mathbf{L}_{\rm total}$ is conserved, the precession of the disc is in the opposite sense to the precession of the planet \citep[e.g.][]{Larwood:1996fk}.

The shared precession time-scale between the inner and outer disc suggest that the disc is precessing as one solid body, with a warp between the inner and outer disc sections demonstrated by the absolute offset in twist between the two regions. We expect global precession of the disc in the case where the disc can communicate the precession on a time-scale faster than the precession occurs \citep{Papaloizou:1995fk}. With the warp propagating at half the sound speed, the communication time-scale is given by 
\begin{align}
t_s=\int_{x_{\rm in}}^{x_{\rm out}} \frac{2}{c_{\rm s,0} (R/R_0)^{-q}} dR, \label{equation:ts}
\end{align}
where $x_{\rm in}$ and $x_{\rm out}$ denote the radial integration limits. We estimate that communication across the disc with $x_{\rm in} = 0.1$~au, $x_{\rm out} =20$~au and $q=0.25$ should take about 16 planet orbits.

To confirm that the disc is precessing at the expected rate, we consider the twist profile in Fig.~\ref{fig:matching_XGPap2013}, right panel and use Equation~\ref{equation:tp}. The precession time-scale evolves throughout the simulation: in the first 25 orbits $t_{\rm prec, \gamma} \sim$580 orbits; in the last 25 orbits $t_{\rm prec, \gamma} \sim$2950 orbits. We can compare this to the precession time estimated from a comparison of the angular momentum and torque on the inner disc \citep{Larwood:1996fk},
\begin{align}
\omega_{\rm prec}= - \frac{3Gm}{4a^3} \frac{ \int_{x_{\rm in}}^{x_{\rm out}} \Sigma R^3 dR}{ \int_{x_{\rm in}}^{x_{\rm out}} \Sigma \Omega_k R^3 dR} \cos \beta'. \label{equation:prec_time}
\end{align}
Here $\beta'$ represents the angle between the planet orbital plane and the inner disc plane, taken as $\sim 7.5^{\circ}$ (i.e. the difference between the black and blue line, Fig.~\ref{fig:matching_XGPap2013}). From $t_{\rm prec}~=~2\pi/ \omega_{\rm prec}$ and with $x_{\rm in} = R_{\rm in}$, $x_{\rm out} = a$, this predicts a precession time-scale of $\sim$400 planet orbits. The discrepancy between the precession time-scales measured with Equation~\ref{equation:prec_time} and directly from the simulation is due to the evolution in the surface density profile, which is not taken into account in Equation~\ref{equation:prec_time}. Over the course of the simulations, the surface density evolves due to accretion on to the star and the planet (e.g. see Fig.~\ref{fig:sigma_profiles}). The closer agreement at earlier times confirms this, as the disc has not yet evolved significantly and the surface density profile is closer to the assumed analytical form (Equation~\ref{equation:sigma}).

As we measure $\beta$ relative to $\mathbf{L}_{\rm total}$, direct comparison of our results in Fig.~\ref{fig:matching_XGPap2013} to fig. 11 of \citet{Xiang-Gruess:2013fg} is not possible. We thus repeat our analysis, measuring $\beta$ relative to the $z$ axis as in their work. Here the disc reaches a final inclination of $\approx$12.5$^{\circ}$, about twice that measured by \citet{Xiang-Gruess:2013fg} for the same initial planet inclination. As our disc is less than half the mass of theirs, this is expected. The relative inclination between the discs reaches a maximum of $\sim4$$^{\circ}$ after 75 orbits, decreasing to $\approx$2.5$^{\circ}$ by the end of our simulation. The precession rate measured from our simulation of $\sim1800$ planet orbits (averaged across 200 orbits) is also in line with the precession rate of $\sim$1000 orbits inferred from their fig. 12.

\subsection{How large should $R_{\rm out}$ be?}
\label{subsection:outeredge}
Previous studies that consider the evolution of a disc in response to an inclined planet \citep[e.g.][]{Bitsch:2013hg,Xiang-Gruess:2013fg,Arzamasskiy:2017ic} have constrained the outer disc radius in their simulations to $\lesssim20$~au. However, most protoplanetary discs are observed to extend to $>100$~au \citep{Andrews:2007pr,Ansdell:2018vi}. Increasing the outer radius increases the communication time-scale in the outer disc significantly, and also may alter the disc response such that the disc may warp rather than tilt as a solid body.

To investigate the effect of $R_{\rm out}$, we repeat the simulation L1 Low with successively larger outer radii. We use the lower resolution version of the simulation shown in Fig.~\ref{fig:pretty} as our initial comparison (L1 Low), with the evolution of the tilt shown in the left panel of Fig.~\ref{fig:outer_edge_resolution_test}. The broad evolution of the tilt is in agreement with the higher resolution version of this simulation, shown in the left panel of Fig.~\ref{fig:matching_XGPap2013}.

For our larger outer radius simulations L2 and L3, we scale the disc mass in each case so that 0.01M$_{\odot}$ is always distributed between 0.1 and 100~au. To maintain a consistent resolution between the simulations we also scale the number of particles, requiring 3.0$\times10^5$ particles for an outer radius of $R_{\rm out}$~=~50~au (L2) and 7.3$\times 10^5$ for $R_{\rm out}$~=~100~au (L3). The only difference between these three simulations is thus how much of the outer disc material is modelled (i.e. the distribution of the angular momentum between the disc components and the planet) and the communication time-scale due to wave propagation across the whole disc.

Fig.~\ref{fig:outer_edge_resolution_test}, middle and right panels shows the tilt evolution for these simulations. We note that as $R_{\rm out}$ increases it takes longer for initial transients to fade, but each of these simulations are run for more than a sound crossing time. In all cases, we find similar behaviour: the planet inclination moves towards $\mathbf{L}_{\rm total}$, and a relative inclination develops between the inner and outer disc. As the tilt is measured from $\mathbf{L}_{\rm total}$, a lower tilt suggests having a larger portion of the total angular momentum in the system --- i.e. when the inclination of the planet is lower than that of the inner and outer disc, the planet has more angular momentum than the total disc. As $R_{\rm out}$ is increased and more of the angular momentum resides with the disc, the tilt of the planet (black line) moves to larger inclinations. We note that as the outer radius increases, $\mathbf{L}_{\rm total}$ moves but the planet orientation with respect to the disc does not. This means that the inclination of the planet $\beta_{\rm planet}$ changes with increasing $R_{\rm out}$, but the relative inclination between the disc and planet is the same (i.e. $\beta_{\rm planet} + \beta_{\rm disc}$ = 20$^{\circ}$).

Within the course of these simulations the relative inclination between the inner and outer disc reaches a constant value, and the discs evolve together. In the $R_{\rm out}=50$~au case, a constant offset of $\approx$1.0$^{\circ}$ develops within about 100 orbits, and in the $R_{\rm out}=100$~au case we see an offset of $\approx$2.0$^{\circ}$ by $\sim$100 orbits. Beyond 350 planet orbits, accretion on to the star and planet has degraded the resolution in the region interior to the planet's orbit and we do not consider the quantitative evolution.

The contrast of the behaviour between these extended discs and the smaller disc simulation is mainly dictated by the ratio of the angular momenta. In the case where the outer disc is more extended, it has a larger angular momentum than the planet and thus does not respond to the influence of the planet as strongly. In all cases the ratio between the planet and inner disc angular momentum is constant ($L_{\rm p}/L_{\rm int} = 20$), so the inner disc always tilts in response to the planet. In the simulation with the largest outer radius, this therefore drives a stronger relative inclination between the inner and outer disc. This suggests that the outer disc should be modelled well past the location of the planet orbit such that the angular momentum of the disc is larger than the planet. As the angular momentum of the outer disc is greater than the planet for both $R_{\rm out}=50$~au and 100~au, the behaviour of the disc is similar. In our subsequent simulations, we thus use an outer radius of 50~au (or 10 times the semi-major axis of the planet).

We find that our resolution decreases quickly during the course of our simulation when using $N=3.0\times 10^5$ in contrast to \citet{Xiang-Gruess:2013fg}. This is due to our different accretion prescriptions. At the location of the planet, our resolution starts at $\langle h \rangle /H = 0.6$ but degrades to 1.7 after 200 planet orbits. This leads to poor resolution of the gas interior to the planets orbit and is exacerbated by the gap carved by the planet. This can be seen in the higher resolution version of this simulation in Fig.~\ref{fig:pretty}, where the material inside 5~au not well resolved.  This means that the tilt measured in the inner disc is not as accurate as that measured in the outer disc, where the disc is remains well resolved. However, our comparison with the higher resolution simulation L1 Med, also included in the left panel of Fig.~\ref{fig:outer_edge_resolution_test}, confirms the behaviour of the disc at low resolution is broadly accurate. We do not consider the evolution of the disc at times greater than $t =350$ orbits with $N=3.0\times 10^5$ particles, as we no longer consider the inner disc to be well resolved.

As increasing $R_{\rm out}$ also changes the sound crossing time-scale in the total disc, each of the simulations in Fig.~\ref{fig:outer_edge_resolution_test} have been run to a different number of sound crossing times. In the largest disc case with $t_{\rm s} \sim160$ orbits, the simulation has run for roughly two crossings. In the case with $R_{\rm out} = 50$~au the disc has run for almost seven sound-crossings (with $t_{\rm s} \sim 50$ orbits), and we find a roughly constant relative inclination has developed after about two of these. As these two larger outer disc simulations have a similar distribution of angular momentum and a similar behaviour, we consider them to be directly comparable. Thus the results from $R_{\rm out}=50$~au suggests that the relative inclination between the inner and outer disc in the $R_{\rm out} = 100$~au simulation is not likely to change if the simulation were to run for longer. Additionally, the similarities between these two simulations implies that the behaviour of the disc may be accurately modelled with an outer radius as small as $R_{\rm out} \gtrsim50$~au. These simulations demonstrate that the evolution of the whole disc is dictated by the size of the outer disc modelled. This is particularly important in the inner region, where a physically motivated $R_{\rm out}$ is critical to determining the relative inclination between the inner and outer disc.

\subsection{Driving warps}
\label{subsection:warping}
Our final set of simulations (L Med with a corresponding convergence test L Low) consider three planet masses and three initial planet inclinations to investigate what parameters may be expected to drive a warp in a protoplanetary disc. \citet{Arzamasskiy:2017ic} suggest that an inclination of $3.0 \times H/R$ drives the largest disc tilt over a fixed number of orbits, so we use inclinations of $i=2.15^{\circ}, 4.30^{\circ}$ and $12.4^{\circ}$ (equivalent to 0.5, 1.0 and 3.0$\times H/R$ for these simulations). The total disc mass is held constant as before, with 0.01M$_{\odot}$ between 0.1 and 100~au but we set $R_{\rm out}$ = 50~au. With planet masses of 0.1, 1.0 and 5$\times$ the $m_{\rm thermal}$, this corresponds to a ratio between the angular momentum of the planet and the gas interior to the planet between 0.4 and 20. We conduct these simulations with $N$=3.4$\times 10^6$ particles, reaching the same resolution as the set of G Med simulations. Across these parameter values, the disc behaviour is a less dramatic version of Fig.~\ref{fig:pretty}, as we consider lower planet inclinations here.

\begin{figure}
	\includegraphics[width=\columnwidth]{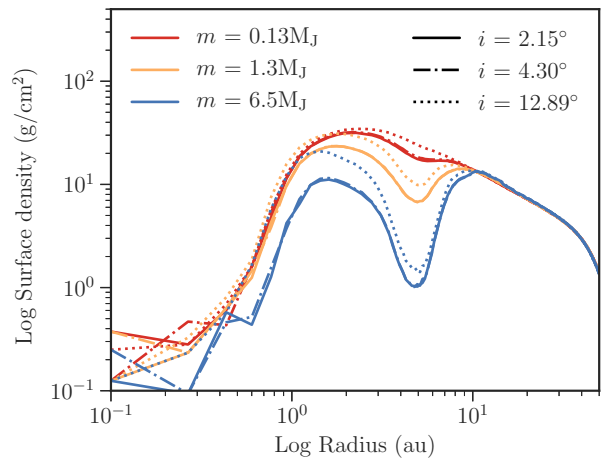}
    \caption{Surface density profiles for the L Med simulations after 100 orbits for $0.13 \rm M_J$ (red), $1.3 \rm M_J$ (orange), $6.5 \rm M_J$ (blue) planet inclined at 2.15$^{\circ}$ (solid line), 4.30$^{\circ}$ (dot-dashed line) and 12.89$^{\circ}$ (dotted). As expected, forming a gap is easier at lower inclinations with more massive planets.\label{fig:sigma_profiles}}
\end{figure}

Fig.~\ref{fig:sigma_profiles} shows the surface density profiles for these simulations after 100 orbits. Evidence of the planet carving a gap is seen for our highest mass planet ($5 m_{\rm thermal}$, shown in blue) and the lower inclination simulations with $m_{\rm thermal}$ planet (shown in orange). We find that a planet that should be able to carve a gap struggles to do so when on an orbit that is strongly inclined to the disc (e.g. by comparing the dotted and solid lines for each planet mass). We do not find significant deviations to the surface density profile for planets that are smaller than $m_{\rm thermal}$, consistent with \citet{Chametla:2017ke}.

\begin{figure*}
	\includegraphics[width=\textwidth]{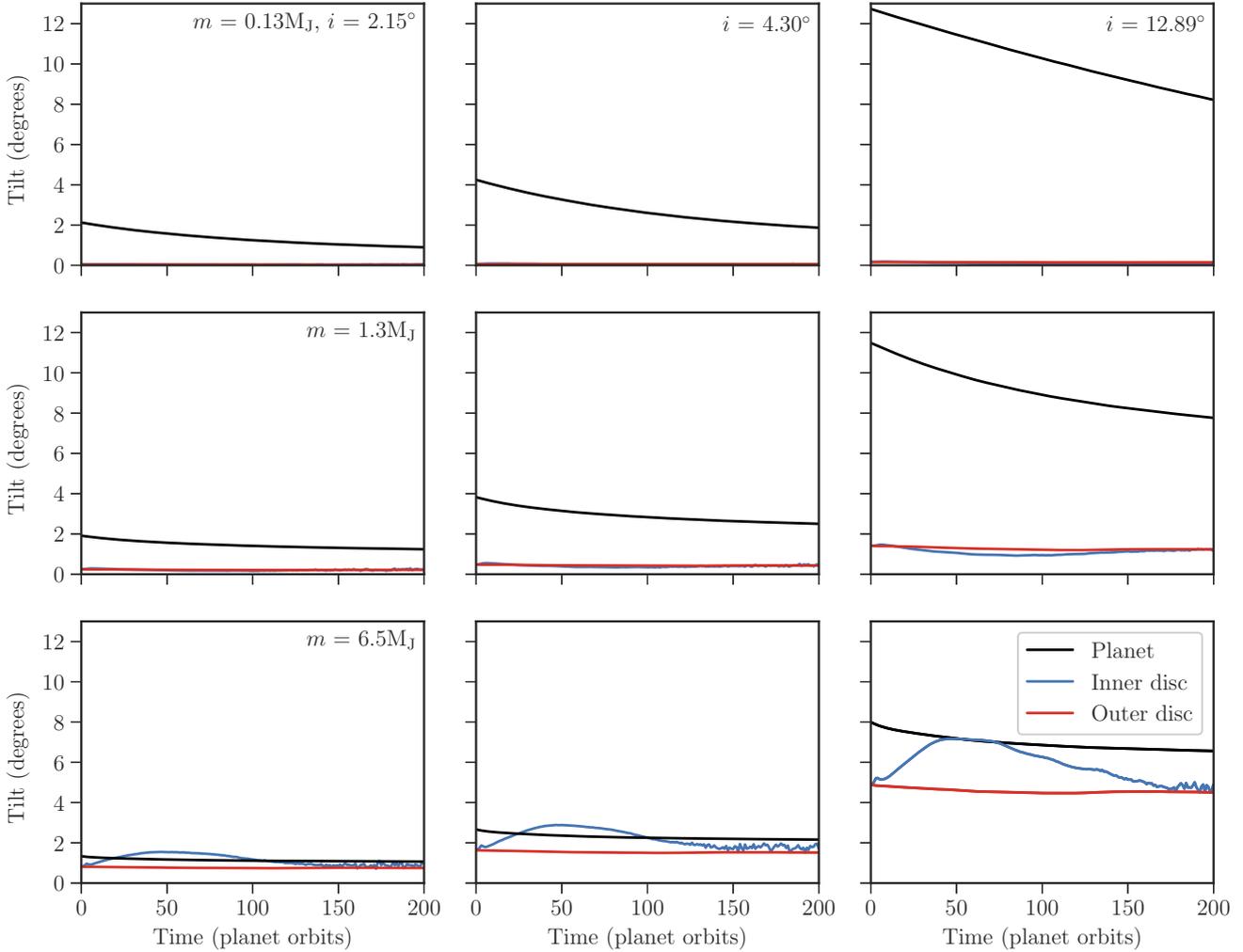}
    \caption{Averaged tilt of the disc interior to the planet orbit (blue), exterior to the planet orbit (red) and the planet orbit (black) relative to the total angular momentum vector of the system. In all cases, the angular momentum of the outer disc is greater than the planet. While the relative inclination between the inner and outer disc is largest for the higher mass planet with a larger inclination orbit (see bottom right panel), moderate relative inclinations are also generated for the lower planet inclinations used.\label{fig:tilt_composite}}
\end{figure*}

Fig.~\ref{fig:tilt_composite} displays the tilt of the gas interior and exterior to the planet orbit for these simulations over 200 planet orbits. The low mass planets (upper and middle panels) have little to no affect on the disc tilt, irrespective of their inclination. As the planet mass increases to $5m_{\rm thermal}$ and the planet is able to easily carve a gap, relative inclinations develop most clearly for the most massive planet at the highest inclination (of $\approx 2^{\circ}$ after 50 orbits). At the highest planet mass (lower panel), a gap is easily carved and the inner and outer discs are misaligned for all three cases. Towards the end of these simulations, the inner and outer disc reach the same relative inclination in all cases.

\begin{figure*}
	\includegraphics[width=\textwidth]{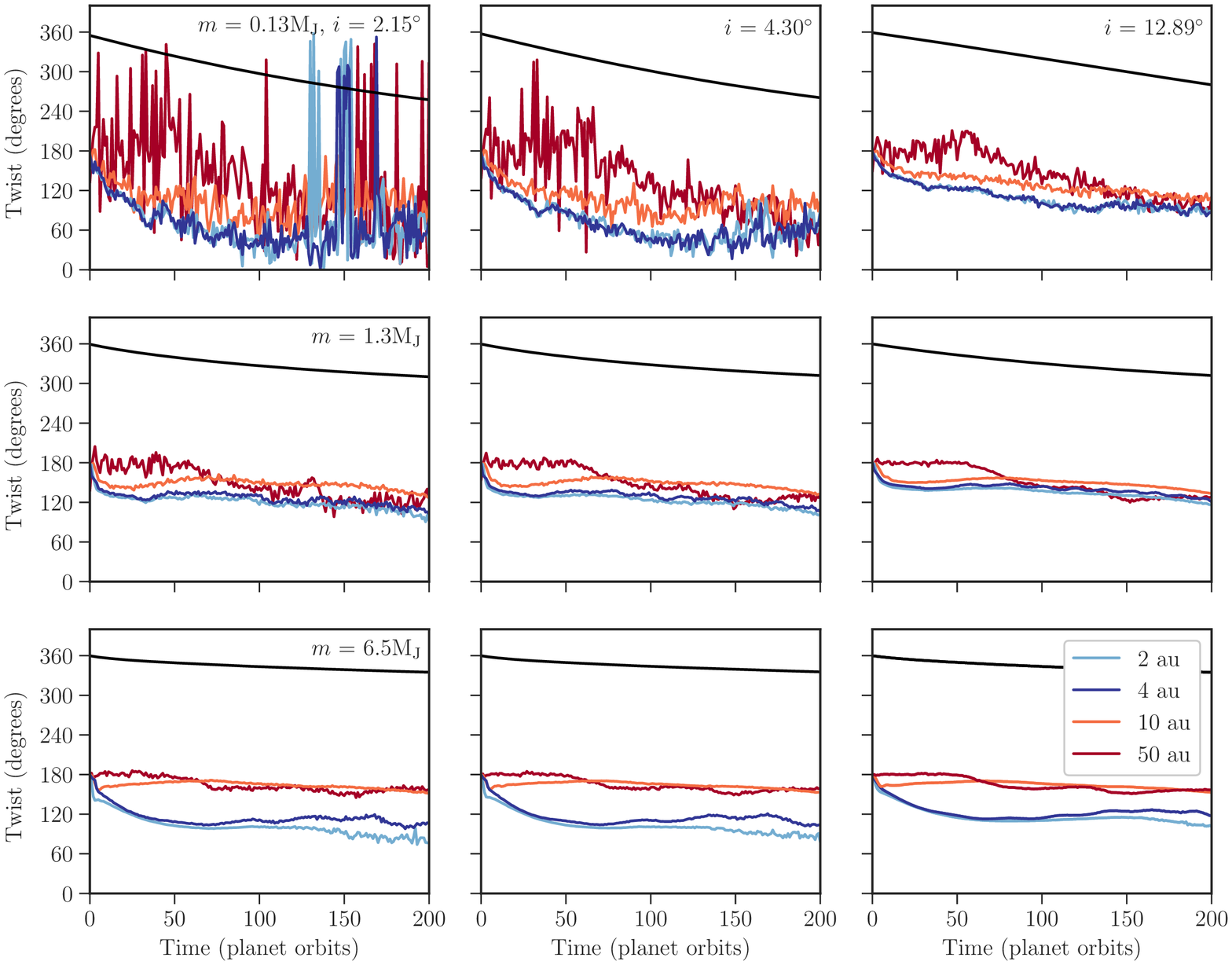}
    \caption{Twist of the gas interior to the planet orbit (2~au in light blue, 4~au in dark blue), exterior to the planet orbit (10~au in orange, 50~au in red) and the planet (black) measured relative to the total angular momentum of the system. The inner and outer edge of the disc outside of the planet orbit twist at different rates, suggesting differential precession between the inner and outer disc although each is precessing as a solid body. \label{fig:twist_composite}}
\end{figure*}

The twist of the inner disc (at 2~au in light blue, 4~au in dark blue), the outer disc (10~au in orange, 50~au in red) and the planet (black) for these simulations is shown in Fig.~\ref{fig:twist_composite}. As before, in this representation the same twist and rate of twist at different radii indicates global, solid body precession between those radii. In all cases here, we find that the gas at 2~au and 4~au has broadly the same absolute twist and is precessing at the same rate, indicating that the gas interior to the planet is precessing as a tilted solid body.

The gas at 10~au and 50~au is precessing with the same rate, indicating that the outer disc is also precessing as a solid body. In the two lower planet mass cases, the inner and outer disc precess with a shared, common rate such that the whole disc is precessing as a solid body. However, in the most massive planet case the inner disc  precesses faster than the outer disc. In our simulations where the planet is massive enough to carve a gap in the disc, we measure differential precession between the inner and outer disc, each precessing as a solid body. We can confirm this behaviour in the outer disc by examining the time-scales: from Equation~\ref{equation:ts} (with $R_{\rm in} = 0.1$~au and $R_{\rm out} = 50$~au), $t_s \gtrsim 50$ planet orbits and the precession time-scale $t_{\rm prec} \gtrsim 600$ orbits (e.g. for $6.5 \rm M_J$ at $12.89^{\circ}$). As the precession time-scale is larger than the communication time-scale, the solid body precession observed by the outer disc in our simulations is expected.

\subsection{What drives the largest warp?}
\label{subsection:warp_strength}
In the context of discs with asymmetric features, we are interested in what combination of planet mass and inclination will induce the largest difference between the inner and outer disc. There are a number of geometric arrangements that can cast a shadow on the outer disc. They may be generally represented by a large warp between the inner and outer disc, caused by a change in tilt and twist across the orbit of the planet. As an example, a disc with a relative tilt between the inner and outer disc will cause shadowing on to the outer disc and the magnitude of the tilt will dictate the amplitude of any brightness variations in the outer disc. In this case, the twist will modulate where such a shadow appears, but not its amplitude. However, in the case where both the inner and outer disc are inclined and there is no relative inclination between the two, the twist angle will determine the strength of the shadow and the location. To determine which parameters lead to the strongest warp, we thus must consider both the tilt and twist. We introduce $\Delta$ as the difference between the unit angular momentum vectors of the inner ($\boldsymbol{\hat{\ell}}_{\rm inner}$) and outer ($\boldsymbol{\hat{\ell}}_{\rm outer}$) discs such that
\begin{align}
\Delta= | \boldsymbol{\hat{\ell}}_{\rm inner} - \boldsymbol{\hat{\ell}}_{\rm outer} |.  \label{equation:warp_strength}
\end{align}
With this representation, $\Delta$ is zero when the angular momentum vectors of the inner and outer disc are aligned (i.e. there is no relative tilt or twist) and a maximum when they are anti-aligned. For our purposes, an observable feature is likely to be driven by a strong change in the tilt and twist corresponding to a large value of $\Delta$. To calculate each vector component $\boldsymbol{\hat{\ell}}_{\rm inner}$ and $\boldsymbol{\hat{\ell}}_{\rm outer}$, we calculate the angular momentum vectors averaged across the gas interior to and exterior to the planet orbit.

\begin{figure}
	\includegraphics[width=\columnwidth]{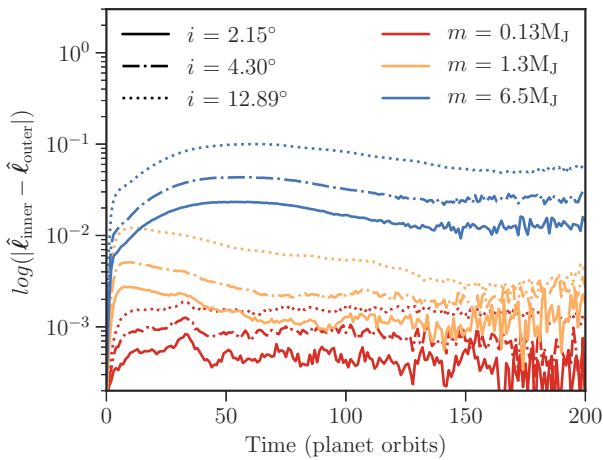}
    \caption{The difference between the inner and outer disc $\Delta$ (with a log-scale for clarity), where larger values represent the strongest change in the tilt and twist across the orbit of the planet and are most likely to produce observable features. We find significant warps can be generated with massive planets with high and moderate inclinations. Here the planet masses are denoted by $0.13 \rm M_J$ (red), $1.3 \rm M_J$ (orange), $6.5 \rm M_J$ (blue) and inclinations by 2.15$^{\circ}$ (solid line), 4.30$^{\circ}$ (dot-dashed line) and 12.89$^{\circ}$ (dotted).\label{fig:warp_strength}}
\end{figure}

Fig.~\ref{fig:warp_strength} shows $\Delta$ for the L Med simulations, corresponding to the tilts and twists in Figs~\ref{fig:tilt_composite} and \ref{fig:twist_composite}, respectively. As expected, the largest warps are driven by massive planets ($6.5 \rm M_J$) on high inclination orbits ($12.89^{\circ}$), where the most significant warp corresponds to a relative twist of 60$^{\circ}$ established after 50 orbits (see the bottom right panels of Figs~\ref{fig:tilt_composite}~and~\ref{fig:twist_composite}). The three strongest warps are driven by the most massive planet, suggesting that the mass of the planet is more important than the inclination for driving potentially observable features. From the right, upper and right, middle panels of Fig.~\ref{fig:tilt_composite} and \ref{fig:twist_composite} these massive planets generate a warp across the disc but no relative tilts at the end of the simulation --- in these scenarios, the differential precession between the inner and outer disc would be responsible for any shadowing on the outer disc rather than a strong relative tilt. The more massive the planet, the cleaner the separation between the inner and outer disc and thus the greater the difference in twist that can be driven.

\section{Discussion}
\label{section:disc}
\subsection{Dependence on disc and planet properties}
We have discussed the simulations in this work focussing on the relationship between the angular momentum of the disc and the planet, and we have demonstrated that altering this ratio can lead to different behaviour of the disc (Fig.~\ref{fig:outer_edge_resolution_test}). However, such a comparison wraps together the absolute and relative relationships between the disc mass, disc size, planet mass, planet inclination and planet orbit in a complicated fashion. In the case that a precessing inner disc does form, the precession time-scale is additionally related to the planet properties. While the precession time-scale does not depend on the mass of the disc (rather, it depends on the radial distribution of the mass, as the $\Sigma_0$ dependence drops out of Equation~\ref{equation:prec_time}), the mass of the disc relative to the planet mass does dictate whether an inner disc can be formed in order to precess. Finally, when we consider the warps that can be driven by an inclined planets, we have demonstrated that similar warps can be made using a combination of different parameters in Fig.~\ref{fig:warp_strength} --- this comparison additionally does not include the effect of the total disc mass. There are thus significant degeneracies between the parameters used in these simulations if compared directly to observations.

\subsection{Time-scales}
We have already confirmed that the sound crossing time-scale is much faster than the precession time-scale when $R_{\rm out} = 50$~au, such that the inclined planet causes the disc to precess as a solid body. By comparing the time-scale for inclination damping with the gap-opening time-scale, we can further show that massive planets will open a gap before their inclination damps. Opening a gap reduces the planet-disc interactions such that the planet may maintain its misalignment. The viscous time-scale
\begin{align}
t_{\nu} = \frac{R^2}{\nu} = \left( \frac{H}{R} \right)^{-2} \frac{1}{\alpha \Omega_k},
\end{align}
is evaluated at the location of the planet to be $\sim 2.8 \times 10^4$ planet orbits for $\alpha=0.001$. Assuming that the planet is massive enough to carve a gap of radial width $\sim H$, the gap opening time-scale is given by \citep{Armitage:2013bg}
\begin{align}
t_{\rm gap} =  \left( \frac{H}{R} \right)^{2} t_{\nu}.
\end{align}
For our estimate of the viscous time-scale at the location of the planet, this corresponds to $\sim 160$ planet orbits.
To estimate the inclination damping of the planet, we can consider the time-scale for a low mass planet \citep{Tanaka:2004od}
\begin{align}
t_{\rm inc} &= \Omega_k^{-1} \left( \frac{H}{R} \right)^4 \left( \frac{m}{M_*} \right)^{-1} \left( \frac{\Sigma R^2}{M_*}\right)^{-1}.
\end{align}
Although this relation is strictly only true for $m \ll m_{\rm thermal}$, we can use it as an lower limit for the inclination damping for a planet massive enough to carve a gap. Taking the disc properties we have used for these simulations ($\rm M_{disc} =0.01\rm M_{\odot}$ between 0.1 and 100~au, $H/R =0.075$ at the location of the planet in our locally isothermal simulations) and a mass ratio of $m/\rm M_*\approx10^{-4}$ implies an inclination damping time-scale of $\sim620$ planet orbits. For a gap carving planet we thus anticipate that the inclination damping time-scale will be longer than $\sim 600$ planet orbits, as suggested by our simulations. This time-scale is longer than the time-scale required to carve a gap, such that $t_{\rm gap} < t_{\rm inc} \ll t_{\nu}$. We thus predict that as the planet inclination damps, the planet carves a gap and the inner disc warps relative to the outer disc. This is consistent with the results of our simulations that consider $m \gtrsim m_{\rm thermal}$.

\subsection{Comparison to \citet{Arzamasskiy:2017ic}}
Although our code test of the radial migration and inclination damping rates shown in Section~\ref{section:linear} are consistent with \citet{Arzamasskiy:2017ic}, our subsequent results with higher mass planets are not. In the simulations with higher planet masses, the case with $1.3 \rm M_J$ inclined at 12.89$^{\circ}$ is closest in parameters to the simulation that they used to generate a polarized light intensity image. While we find evidence of a warp in our simulations (Fig.~\ref{fig:warp_strength}, orange dotted line), the close comparison of the co-planar and inclined cases in fig. 9 of \citet{Arzamasskiy:2017ic} suggest that they did not find the same. 

Beyond the different numerical methods we have used, there are a numerous distinctions between the simulations presented here and in \citet{Arzamasskiy:2017ic} that likely contribute to the difference. First, in our simulations, the planet is free to migrate while \citet{Arzamasskiy:2017ic} use a fixed orbit. Second, the outer disc edge in the simulation presented by \citet{Arzamasskiy:2017ic} is three times larger than the orbit of the planet, while in ours it is ten times larger. In Fig.~\ref{fig:outer_edge_resolution_test} we demonstrated that the relative inclination between the inner and outer disc depends on the angular momentum ratios considered, with larger relative inclinations measured in discs that are simulated with a larger radial extent. It is thus likely that our larger outer radius contributed to the warp measured here. Third, although the simulation conducted by \citet{Arzamasskiy:2017ic} has run for longer than the sound crossing time, it was not long enough for the behaviour of the disc to be fully established (they quote 15 planet orbits). As shown by \citet{Xiang-Gruess:2013fg} and \citet{Bitsch:2013hg}, hundreds of orbits are required to reveal the evolution of the disc when influenced by a misaligned planet. In addition to this, our simulations with the larger outer radius increase the sound crossing time, thus requiring even longer to observe significant evolution of the disc.

Finally, \citet{Arzamasskiy:2017ic} considered the evolution of an inviscid disc while our simulations include viscosity. The relative inclinations that are measured in our simulations are consistent with previous results that included viscosity using SPH by \citet{Xiang-Gruess:2013fg} and using the grid code \textsc{NIRVANA} \citet{Bitsch:2013hg}, taking into account the planet masses that we have used here. In a warped disc the radial pressure gradient arising between misaligned neighbouring rings oscillates around the orbit, inducing epicyclic motion. In near-Keplerian discs, this motion resonates with the orbital forcing. In the absence of viscosity this leads to the launching of a bending wave which propagates through the disc with a wave-speed $\sim c_{\rm s}/2$ (\citealt{Papaloizou:1995pn}; see \citealt{Nixon:2016nh} for a review of warped disc physics). Introducing a small viscosity (such that $\alpha < H/R$) causes this wave to damp as it propagates on a time-scale $t_{\rm damp} = 1/\alpha\Omega_k$ \citep{lubow_ogilvie_2000}. Thus the viscosity effects the location and rate of dissipation of the motions induced in the disc by the planet. Including dissipation through a viscosity \citep[e.g.][]{Bate:2000fk,king_et_al_2013} allows the evolution of the angular momentum vector of the planet and that of the disc towards the total angular momentum vector \citep{King:2005qy}.


\subsection{Strongly misaligned discs}
We caution that although there are observations of strongly misaligned planets \citep[measured obliquities of hot Jupiters may easily be up to $\sim$30$^{\circ}$ measured with respect to the stellar spin, e.g.][]{Triaud:2010vj,Albrecht:2012ih}, these are unlikely to be responsible for discs with large misalignments ($\gtrsim$45$^{\circ}$, as observed in HD 100453 and HD 100564). In the case that a planet does have strong misalignments the interaction between the disc and planet are reduced to the brief moment that the planet `punches' through the disc twice per orbit. Simulations by \citet{Xiang-Gruess:2013fg} demonstrated that in this case, the interactions are enough to damp the inclination of the planet but not enough to drive significant evolution of the disc. Instead, the planet inclination decreases until it is small enough that it can produce a gap. In this interpretation, disc misalignments appear to be unable to be driven by a planet with a large inclination. For those discs that are strongly misaligned, \citet{Bate:2018ls} has shown that separate infall events with independent angular momenta can generate two misaligned disc sections (see their fig. 2).

\subsection{Limitations}
As discussed in Section~\ref{subsection:outeredge}, a major limitation with simulations of protoplanetary discs is the size and duration of the simulation. For SPH simulations in particular, the resolution of our simulations may be reduced close to the star and planet as they accrete particles. This limits the duration of our simulations to $\sim$200 orbits, after which we do not consider the inner disc to be well resolved --- although the precession time-scale of these discs is much longer than this.


These simulations are also limited by only including a single planet on a circular orbit in an isothermal disc. In the event that this planet is misaligned due to scattering, a second planet of similar mass will reside in the disc (when not ejected) and it is expected that the orbits of both will have a non-zero eccentricity. Whether or not the inclination damps for highly eccentric planets depends on the properties of the disc; highly eccentric massive planets in a radiative disc can drive oscillations in the inclination \citep{Bitsch:2011ne} while in an isothermal disc the inclination damps \citep{Bitsch:2013hg}. Additionally, including radiation may drive outwards migration \citep{Bitsch:2011ne}.


\section{Conclusions}
\label{section:concs}
In this paper we use three-dimensional numerical simulations to investigate the response of a protoplanetary disc to a misaligned planet. We begin by studying a planet in the linear mass regime where the planet has a negligible effect on the disc, demonstrating that our numerical code of choice \textsc{Phantom} is able to accurately model the motion of a planet in this regime. We then consider more massive planets that are able to affect the structure of the disc.

We confirm that a planet massive enough to carve a gap is able to separate the disc into two distinct discs, where the inclination of the planet drives a relative tilt and twist between the inner and outer disc. We demonstrate that altering the outer radius of the total disc alters the magnitude of the warp that develops. This is due to the angular momentum balance between the inner disc, outer disc and the planet. In the case that the outer radius is much larger than the planet orbit, the disc holds a larger fraction of the total angular momentum than the planet and does not respond strongly to the influence of the planet. The planet is still able to tilt the disc interior to its orbit (as this has less angular momentum than the planet), and so a larger relative tilt develops between the inner and outer discs than in simulations with a smaller outer radius. These results demonstrate that the outer radius is critical to the evolution of the warp in the inner regions of discs warped by misaligned planets.

We examine the warp generated by planets with different masses and inclinations. To estimate which of these causes the largest (and most observationally relevant) warp we consider the difference between the unit angular momentum vectors of the inner and outer disc in each case. Importantly, this measure takes into account both the tilt and twist between each of the discs, either of which may lead to shadowing on the outer disc. We find that the mass of the planet is more relevant for generating large warps than the inclination (in the case that the planet is able to carve a gap). This suggests that the ability to carve a gap and separate the disc is the most important feature to generate a warp, with massive planets at moderate inclinations ($i \lesssim 3 \times H/R$) potentially driving significant warps.


\section*{Acknowledgements}
We thank the anonymous referee, Alessia Franchini and Cassandra Hall for useful comments and discussions. This project has received funding from the European Research Council (ERC) under the European Union's Horizon 2020 research and innovation programme (grant agreement No 681601). RGM acknowledges support from NASA through grant NNX17AB96G. CN is supported by the Science and Technology Facilities Council (STFC) (grant number ST/M005917/1). Results described here were obtained using the Darwin Supercomputer of the University of Cambridge High Performance Computing Service as part of the DiRAC facility jointly funded by STFC and the Large Facilities Capital Fund of BIS. We used \textsc{SPLASH} \citep{Price:2007kx} for Fig.~\ref{fig:pretty}.





\bibliographystyle{mnras}
\bibliography{/Users/rln12/Documents/my_absolute_latex/bib/master} 

\begin{thebibliography}{}
\makeatletter
\relax
\def\mn@urlcharsother{\let\do\@makeother \do\$\do\&\do\#\do\^\do\_\do\%\do\~}
\def\mn@doi{\begingroup\mn@urlcharsother \@ifnextchar [ {\mn@doi@}
  {\mn@doi@[]}}
\def\mn@doi@[#1]#2{\def\@tempa{#1}\ifx\@tempa\@empty \href
  {http://dx.doi.org/#2} {doi:#2}\else \href {http://dx.doi.org/#2} {#1}\fi
  \endgroup}
\def\mn@eprint#1#2{\mn@eprint@#1:#2::\@nil}
\def\mn@eprint@arXiv#1{\href {http://arxiv.org/abs/#1} {{\tt arXiv:#1}}}
\def\mn@eprint@dblp#1{\href {http://dblp.uni-trier.de/rec/bibtex/#1.xml}
  {dblp:#1}}
\def\mn@eprint@#1:#2:#3:#4\@nil{\def\@tempa {#1}\def\@tempb {#2}\def\@tempc
  {#3}\ifx \@tempc \@empty \let \@tempc \@tempb \let \@tempb \@tempa \fi \ifx
  \@tempb \@empty \def\@tempb {arXiv}\fi \@ifundefined
  {mn@eprint@\@tempb}{\@tempb:\@tempc}{\expandafter \expandafter \csname
  mn@eprint@\@tempb\endcsname \expandafter{\@tempc}}}

\bibitem[\protect\citeauthoryear{{ALMA Partnership} et~al.,}{{ALMA Partnership}
  et~al.}{2015}]{ALMA:2015ng}
{ALMA Partnership} et~al., 2015, \mn@doi [\apjl] {10.1088/2041-8205/808/1/L3},
  \href {http://adsabs.harvard.edu/abs/2015ApJ...808L...3A} {808, L3}

\bibitem[\protect\citeauthoryear{{Acke} \& {van den Ancker}}{{Acke} \& {van den
  Ancker}}{2006}]{Acke:2006vs}
{Acke} B.,  {van den Ancker} M.~E.,  2006, \mn@doi [\aap]
  {10.1051/0004-6361:20054330}, \href
  {http://adsabs.harvard.edu/abs/2006A%26A...449..267A} {449, 267}

\bibitem[\protect\citeauthoryear{{Albrecht} et~al.,}{{Albrecht}
  et~al.}{2012}]{Albrecht:2012ih}
{Albrecht} S.,  et~al., 2012, \mn@doi [\apj] {10.1088/0004-637X/757/1/18},
  \href {http://ukads.nottingham.ac.uk/abs/2012ApJ...757...18A} {757, 18}

\bibitem[\protect\citeauthoryear{{Andrews} \& {Williams}}{{Andrews} \&
  {Williams}}{2007}]{Andrews:2007pr}
{Andrews} S.~M.,  {Williams} J.~P.,  2007, \mn@doi [\apj] {10.1086/511741},
  \href {http://ukads.nottingham.ac.uk/abs/2007ApJ...659..705A} {659, 705}

\bibitem[\protect\citeauthoryear{{Andrews} et~al.,}{{Andrews}
  et~al.}{2016}]{Andrews:2016bw}
{Andrews} S.~M.,  et~al., 2016, \mn@doi [\apjl] {10.3847/2041-8205/820/2/L40},
  \href {http://adsabs.harvard.edu/abs/2016ApJ...820L..40A} {820, L40}

\bibitem[\protect\citeauthoryear{{Ansdell} et~al.,}{{Ansdell}
  et~al.}{2016}]{Ansdell:2016gh}
{Ansdell} M.,  et~al., 2016, \mn@doi [\apj] {10.3847/0004-637X/828/1/46}, \href
  {http://adsabs.harvard.edu/abs/2016ApJ...828...46A} {828, 46}

\bibitem[\protect\citeauthoryear{{Ansdell} et~al.,}{{Ansdell}
  et~al.}{2018}]{Ansdell:2018vi}
{Ansdell} M.,  et~al., 2018, preprint, \href
  {http://ukads.nottingham.ac.uk/abs/2018arXiv180305923A} {} (\mn@eprint
  {arXiv} {1803.05923})

\bibitem[\protect\citeauthoryear{{Armitage}}{{Armitage}}{2013}]{Armitage:2013bg}
{Armitage} P.~J.,  2013, {Astrophysics of Planet Formation}

\bibitem[\protect\citeauthoryear{{Arulanantham} et~al.,}{{Arulanantham}
  et~al.}{2018}]{Arulanantham:2018og}
{Arulanantham} N.,  et~al., 2018, \mn@doi [\apj] {10.3847/1538-4357/aaaf65},
  \href {http://adsabs.harvard.edu/abs/2018ApJ...855...98A} {855, 98}

\bibitem[\protect\citeauthoryear{{Arzamasskiy}, {Zhu}  \&
  {Stone}}{{Arzamasskiy} et~al.}{2017}]{Arzamasskiy:2017ic}
{Arzamasskiy} L.,  {Zhu} Z.,   {Stone} J.~M.,  2017, preprint, \href
  {http://ukads.nottingham.ac.uk/abs/2017arXiv171011128A} {} (\mn@eprint
  {arXiv} {1710.11128})

\bibitem[\protect\citeauthoryear{{Ayliffe} \& {Bate}}{{Ayliffe} \&
  {Bate}}{2010}]{Ayliffe:2010ow}
{Ayliffe} B.~A.,  {Bate} M.~R.,  2010, \mn@doi [\mnras]
  {10.1111/j.1365-2966.2010.17221.x}, \href
  {http://ukads.nottingham.ac.uk/abs/2010MNRAS.408..876A} {408, 876}

\bibitem[\protect\citeauthoryear{{Baruteau} et~al.,}{{Baruteau}
  et~al.}{2014}]{Baruteau:2014hb}
{Baruteau} C.,  et~al., 2014, \mn@doi [Protostars and Planets VI]
  {10.2458/azu_uapress_9780816531240-ch029}, \href
  {http://adsabs.harvard.edu/abs/2014prpl.conf..667B} {pp 667--689}

\bibitem[\protect\citeauthoryear{{Bate}}{{Bate}}{2018}]{Bate:2018ls}
{Bate} M.~R.,  2018, \mn@doi [\mnras] {10.1093/mnras/sty169}, \href
  {http://ukads.nottingham.ac.uk/abs/2018MNRAS.475.5618B} {475, 5618}

\bibitem[\protect\citeauthoryear{{Bate}, {Bonnell}  \& {Price}}{{Bate}
  et~al.}{1995}]{Bate:1995fi}
{Bate} M.~R.,  {Bonnell} I.~A.,   {Price} N.~M.,  1995, \mn@doi [\mnras]
  {10.1093/mnras/277.2.362}, \href
  {http://ukads.nottingham.ac.uk/abs/1995MNRAS.277..362B} {277, 362}

\bibitem[\protect\citeauthoryear{{Bate}, {Bonnell}, {Clarke}, {Lubow},
  {Ogilvie}, {Pringle}  \& {Tout}}{{Bate} et~al.}{2000}]{Bate:2000fk}
{Bate} M.~R.,  {Bonnell} I.~A.,  {Clarke} C.~J.,  {Lubow} S.~H.,  {Ogilvie}
  G.~I.,  {Pringle} J.~E.,   {Tout} C.~A.,  2000, \mn@doi [\mnras]
  {10.1046/j.1365-8711.2000.03648.x}, \href
  {http://adsabs.harvard.edu/abs/2000MNRAS.317..773B} {317, 773}

\bibitem[\protect\citeauthoryear{{Bate}, {Lodato}  \& {Pringle}}{{Bate}
  et~al.}{2010}]{Bate:2010nh}
{Bate} M.~R.,  {Lodato} G.,   {Pringle} J.~E.,  2010, \mn@doi [\mnras]
  {10.1111/j.1365-2966.2009.15773.x}, \href
  {http://adsabs.harvard.edu/abs/2010MNRAS.401.1505B} {401, 1505}

\bibitem[\protect\citeauthoryear{{Benisty} et~al.,}{{Benisty}
  et~al.}{2015}]{Benisty:2015na}
{Benisty} M.,  et~al., 2015, \mn@doi [\aap] {10.1051/0004-6361/201526011},
  \href {http://adsabs.harvard.edu/abs/2015A%26A...578L...6B} {578, L6}

\bibitem[\protect\citeauthoryear{{Benisty} et~al.,}{{Benisty}
  et~al.}{2017}]{Benisty:2017kq}
{Benisty} M.,  et~al., 2017, \mn@doi [\aap] {10.1051/0004-6361/201629798},
  \href {http://adsabs.harvard.edu/abs/2017A%26A...597A..42B} {597, A42}

\bibitem[\protect\citeauthoryear{{Bitsch} \& {Kley}}{{Bitsch} \&
  {Kley}}{2011}]{Bitsch:2011ne}
{Bitsch} B.,  {Kley} W.,  2011, \mn@doi [\aap] {10.1051/0004-6361/201016179},
  \href {http://ukads.nottingham.ac.uk/abs/2011A%26A...530A..41B} {530, A41}

\bibitem[\protect\citeauthoryear{{Bitsch}, {Boley}  \& {Kley}}{{Bitsch}
  et~al.}{2013a}]{Bitsch:2013pf}
{Bitsch} B.,  {Boley} A.,   {Kley} W.,  2013a, \mn@doi [\aap]
  {10.1051/0004-6361/201118490}, \href
  {http://ukads.nottingham.ac.uk/abs/2013A%26A...550A..52B} {550, A52}

\bibitem[\protect\citeauthoryear{{Bitsch}, {Crida}, {Libert}  \&
  {Lega}}{{Bitsch} et~al.}{2013b}]{Bitsch:2013hg}
{Bitsch} B.,  {Crida} A.,  {Libert} A.-S.,   {Lega} E.,  2013b, \mn@doi [\aap]
  {10.1051/0004-6361/201220310}, \href
  {http://adsabs.harvard.edu/abs/2013A%26A...555A.124B} {555, A124}

\bibitem[\protect\citeauthoryear{{Booth}, {Walsh}, {Kama}, {Loomis}, {Maud}  \&
  {Juh{\'a}sz}}{{Booth} et~al.}{2018}]{Booth:2018ng}
{Booth} A.~S.,  {Walsh} C.,  {Kama} M.,  {Loomis} R.~A.,  {Maud} L.~T.,
  {Juh{\'a}sz} A.,  2018, \mn@doi [\aap] {10.1051/0004-6361/201731347}, \href
  {http://ukads.nottingham.ac.uk/abs/2018A%26A...611A..16B} {611, A16}

\bibitem[\protect\citeauthoryear{{Burns}}{{Burns}}{1976}]{Burns:1976of}
{Burns} J.~A.,  1976, \mn@doi [American Journal of Physics] {10.1119/1.10237},
  \href {http://ukads.nottingham.ac.uk/abs/1976AmJPh..44..944B} {44, 944}

\bibitem[\protect\citeauthoryear{{Calvet}, {D'Alessio}, {Hartmann}, {Wilner},
  {Walsh}  \& {Sitko}}{{Calvet} et~al.}{2002}]{Calvet:2002vr}
{Calvet} N.,  {D'Alessio} P.,  {Hartmann} L.,  {Wilner} D.,  {Walsh} A.,
  {Sitko} M.,  2002, \mn@doi [\apj] {10.1086/339061}, \href
  {http://adsabs.harvard.edu/abs/2002ApJ...568.1008C} {568, 1008}

\bibitem[\protect\citeauthoryear{{Casassus} et~al.,}{{Casassus}
  et~al.}{2018}]{Casassus:2018te}
{Casassus} S.,  et~al., 2018, \mn@doi [\mnras] {10.1093/mnras/sty894}, \href
  {http://adsabs.harvard.edu/abs/2018MNRAS.tmp..868C} {}

\bibitem[\protect\citeauthoryear{{Chametla}, {S{\'a}nchez-Salcedo}, {Masset}
  \& {Hidalgo-G{\'a}mez}}{{Chametla} et~al.}{2017}]{Chametla:2017ke}
{Chametla} R.~O.,  {S{\'a}nchez-Salcedo} F.~J.,  {Masset} F.~S.,
  {Hidalgo-G{\'a}mez} A.~M.,  2017, \mn@doi [\mnras] {10.1093/mnras/stx817},
  \href {http://ukads.nottingham.ac.uk/abs/2017MNRAS.468.4610C} {468, 4610}

\bibitem[\protect\citeauthoryear{{Cresswell}, {Dirksen}, {Kley}  \&
  {Nelson}}{{Cresswell} et~al.}{2007}]{Cresswell:2007ss}
{Cresswell} P.,  {Dirksen} G.,  {Kley} W.,   {Nelson} R.~P.,  2007, \mn@doi
  [\aap] {10.1051/0004-6361:20077666}, \href
  {http://ukads.nottingham.ac.uk/abs/2007A%26A...473..329C} {473, 329}

\bibitem[\protect\citeauthoryear{{D'Angelo}, {Lubow}  \& {Bate}}{{D'Angelo}
  et~al.}{2006}]{Dangelo:2006ud}
{D'Angelo} G.,  {Lubow} S.~H.,   {Bate} M.~R.,  2006, \mn@doi [\apj]
  {10.1086/508451}, \href
  {http://ukads.nottingham.ac.uk/abs/2006ApJ...652.1698D} {652, 1698}

\bibitem[\protect\citeauthoryear{{Debes} et~al.,}{{Debes}
  et~al.}{2017}]{Debes:2017fk}
{Debes} J.~H.,  et~al., 2017, \mn@doi [\apj] {10.3847/1538-4357/835/2/205},
  \href {http://adsabs.harvard.edu/abs/2017ApJ...835..205D} {835, 205}

\bibitem[\protect\citeauthoryear{{Dipierro}, {Price}, {Laibe}, {Hirsh},
  {Cerioli}  \& {Lodato}}{{Dipierro} et~al.}{2015}]{Dipierro:2015od}
{Dipierro} G.,  {Price} D.,  {Laibe} G.,  {Hirsh} K.,  {Cerioli} A.,   {Lodato}
  G.,  2015, \mn@doi [\mnras] {10.1093/mnrasl/slv105}, \href
  {http://adsabs.harvard.edu/abs/2015MNRAS.453L..73D} {453, L73}

\bibitem[\protect\citeauthoryear{{Dipierro}, {Laibe}, {Price}  \&
  {Lodato}}{{Dipierro} et~al.}{2016}]{Dipierro:2016ss}
{Dipierro} G.,  {Laibe} G.,  {Price} D.~J.,   {Lodato} G.,  2016, \mn@doi
  [\mnras] {10.1093/mnrasl/slw032}, \href
  {http://adsabs.harvard.edu/abs/2016MNRAS.459L...1D} {459, L1}

\bibitem[\protect\citeauthoryear{{Do{\u g}an}, {Nixon}, {King}  \&
  {Price}}{{Do{\u g}an} et~al.}{2015}]{Dogan:2015rt}
{Do{\u g}an} S.,  {Nixon} C.,  {King} A.,   {Price} D.~J.,  2015, \mn@doi
  [\mnras] {10.1093/mnras/stv347}, \href
  {http://adsabs.harvard.edu/abs/2015MNRAS.449.1251D} {449, 1251}

\bibitem[\protect\citeauthoryear{{Facchini}, {Lodato}  \& {Price}}{{Facchini}
  et~al.}{2013}]{facchini_2013}
{Facchini} S.,  {Lodato} G.,   {Price} D.~J.,  2013, \mn@doi [\mnras]
  {10.1093/mnras/stt877}, \href
  {http://adsabs.harvard.edu/abs/2013MNRAS.433.2142F} {433, 2142}

\bibitem[\protect\citeauthoryear{{Flaherty}, {Hughes}, {Rosenfeld}, {Andrews},
  {Chiang}, {Simon}, {Kerzner}  \& {Wilner}}{{Flaherty}
  et~al.}{2015}]{Flaherty:2015is}
{Flaherty} K.~M.,  {Hughes} A.~M.,  {Rosenfeld} K.~A.,  {Andrews} S.~M.,
  {Chiang} E.,  {Simon} J.~B.,  {Kerzner} S.,   {Wilner} D.~J.,  2015, \mn@doi
  [\apj] {10.1088/0004-637X/813/2/99}, \href
  {http://ukads.nottingham.ac.uk/abs/2015ApJ...813...99F} {813, 99}

\bibitem[\protect\citeauthoryear{{Flaherty} et~al.,}{{Flaherty}
  et~al.}{2017}]{Flaherty:2017do}
{Flaherty} K.~M.,  et~al., 2017, \mn@doi [\apj] {10.3847/1538-4357/aa79f9},
  \href {http://ukads.nottingham.ac.uk/abs/2017ApJ...843..150F} {843, 150}

\bibitem[\protect\citeauthoryear{{Flebbe}, {Muenzel}, {Herold}, {Riffert}  \&
  {Ruder}}{{Flebbe} et~al.}{1994}]{Flebbe:1994lr}
{Flebbe} O.,  {Muenzel} S.,  {Herold} H.,  {Riffert} H.,   {Ruder} H.,  1994,
  \mn@doi [\apj] {10.1086/174526}, \href
  {http://adsabs.harvard.edu/abs/1994ApJ...431..754F} {431, 754}

\bibitem[\protect\citeauthoryear{{Goldreich} \& {Sari}}{{Goldreich} \&
  {Sari}}{2003}]{Goldreich:2003er}
{Goldreich} P.,  {Sari} R.,  2003, \mn@doi [\apj] {10.1086/346202}, \href
  {http://ukads.nottingham.ac.uk/abs/2003ApJ...585.1024G} {585, 1024}

\bibitem[\protect\citeauthoryear{{Goldreich} \& {Tremaine}}{{Goldreich} \&
  {Tremaine}}{1979}]{Goldreich:1979ir}
{Goldreich} P.,  {Tremaine} S.,  1979, \mn@doi [\apj] {10.1086/157448}, \href
  {http://ukads.nottingham.ac.uk/abs/1979ApJ...233..857G} {233, 857}

\bibitem[\protect\citeauthoryear{{Goldreich} \& {Tremaine}}{{Goldreich} \&
  {Tremaine}}{1980}]{Goldreich:1908nf}
{Goldreich} P.,  {Tremaine} S.,  1980, \mn@doi [\apj] {10.1086/158356}, \href
  {http://ukads.nottingham.ac.uk/abs/1980ApJ...241..425G} {241, 425}

\bibitem[\protect\citeauthoryear{{Grady}, {Woodgate}, {Heap}, {Bowers}, {Nuth},
  {Herczeg}  \& {Hill}}{{Grady} et~al.}{2005}]{Grady:2005vq}
{Grady} C.~A.,  {Woodgate} B.,  {Heap} S.~R.,  {Bowers} C.,  {Nuth} III J.~A.,
  {Herczeg} G.~J.,   {Hill} H.~G.~M.,  2005, \mn@doi [\apj] {10.1086/426887},
  \href {http://adsabs.harvard.edu/abs/2005ApJ...620..470G} {620, 470}

\bibitem[\protect\citeauthoryear{{Hughes}, {Wilner}, {Andrews}, {Qi}  \&
  {Hogerheijde}}{{Hughes} et~al.}{2011}]{Hughes:2011gw}
{Hughes} A.~M.,  {Wilner} D.~J.,  {Andrews} S.~M.,  {Qi} C.,   {Hogerheijde}
  M.~R.,  2011, \mn@doi [\apj] {10.1088/0004-637X/727/2/85}, \href
  {http://adsabs.harvard.edu/abs/2011ApJ...727...85H} {727, 85}

\bibitem[\protect\citeauthoryear{{Isella}, {P{\'e}rez}, {Carpenter}, {Ricci},
  {Andrews}  \& {Rosenfeld}}{{Isella} et~al.}{2013}]{Isella:2013ng}
{Isella} A.,  {P{\'e}rez} L.~M.,  {Carpenter} J.~M.,  {Ricci} L.,  {Andrews}
  S.,   {Rosenfeld} K.,  2013, \mn@doi [\apj] {10.1088/0004-637X/775/1/30},
  \href {http://adsabs.harvard.edu/abs/2013ApJ...775...30I} {775, 30}

\bibitem[\protect\citeauthoryear{{Kenyon} \& {Hartmann}}{{Kenyon} \&
  {Hartmann}}{1987}]{Kenyon:1987bg}
{Kenyon} S.~J.,  {Hartmann} L.,  1987, \mn@doi [\apj] {10.1086/165866}, \href
  {http://ukads.nottingham.ac.uk/abs/1987ApJ...323..714K} {323, 714}

\bibitem[\protect\citeauthoryear{{King}, {Lubow}, {Ogilvie}  \&
  {Pringle}}{{King} et~al.}{2005}]{King:2005qy}
{King} A.~R.,  {Lubow} S.~H.,  {Ogilvie} G.~I.,   {Pringle} J.~E.,  2005,
  \mn@doi [\mnras] {10.1111/j.1365-2966.2005.09378.x}, \href
  {http://adsabs.harvard.edu/abs/2005MNRAS.363...49K} {363, 49}

\bibitem[\protect\citeauthoryear{{King}, {Livio}, {Lubow}  \& {Pringle}}{{King}
  et~al.}{2013}]{king_et_al_2013}
{King} A.~R.,  {Livio} M.,  {Lubow} S.~H.,   {Pringle} J.~E.,  2013, \mn@doi
  [\mnras] {10.1093/mnras/stt364}, \href
  {http://adsabs.harvard.edu/abs/2013MNRAS.431.2655K} {431, 2655}

\bibitem[\protect\citeauthoryear{{Kley} \& {Nelson}}{{Kley} \&
  {Nelson}}{2012}]{Kley:2012of}
{Kley} W.,  {Nelson} R.~P.,  2012, \mn@doi [\araa]
  {10.1146/annurev-astro-081811-125523}, \href
  {http://adsabs.harvard.edu/abs/2012ARA%26A..50..211K} {50, 211}

\bibitem[\protect\citeauthoryear{{Larwood}, {Nelson}, {Papaloizou}  \&
  {Terquem}}{{Larwood} et~al.}{1996}]{Larwood:1996fk}
{Larwood} J.~D.,  {Nelson} R.~P.,  {Papaloizou} J.~C.~B.,   {Terquem} C.,
  1996, \mnras, \href {http://adsabs.harvard.edu/abs/1996MNRAS.282..597L} {282,
  597}

\bibitem[\protect\citeauthoryear{{Lodato} \& {Price}}{{Lodato} \&
  {Price}}{2010}]{lodato_2010}
{Lodato} G.,  {Price} D.~J.,  2010, \mn@doi [\mnras]
  {10.1111/j.1365-2966.2010.16526.x}, \href
  {http://adsabs.harvard.edu/abs/2010MNRAS.405.1212L} {405, 1212}

\bibitem[\protect\citeauthoryear{{Lubow} \& {Martin}}{{Lubow} \&
  {Martin}}{2016}]{Lubow:2016nw}
{Lubow} S.~H.,  {Martin} R.~G.,  2016, \mn@doi [\apj]
  {10.3847/0004-637X/817/1/30}, \href
  {https://ui.adsabs.harvard.edu/#abs/2016ApJ...817...30L} {817, 30}

\bibitem[\protect\citeauthoryear{{Lubow} \& {Ogilvie}}{{Lubow} \&
  {Ogilvie}}{2000}]{lubow_ogilvie_2000}
{Lubow} S.~H.,  {Ogilvie} G.~I.,  2000, \mn@doi [\apj] {10.1086/309101}, \href
  {http://adsabs.harvard.edu/abs/2000ApJ...538..326L} {538, 326}

\bibitem[\protect\citeauthoryear{{Martin}, {Nixon}, {Armitage}, {Lubow}  \&
  {Price}}{{Martin} et~al.}{2014a}]{Martin:2014aa}
{Martin} R.~G.,  {Nixon} C.,  {Armitage} P.~J.,  {Lubow} S.~H.,   {Price}
  D.~J.,  2014a, \mn@doi [\apjl] {10.1088/2041-8205/790/2/L34}, \href
  {http://adsabs.harvard.edu/abs/2014ApJ...790L..34M} {790, L34}

\bibitem[\protect\citeauthoryear{{Martin}, {Nixon}, {Lubow}, {Armitage},
  {Price}, {Do{\u g}an}  \& {King}}{{Martin} et~al.}{2014b}]{Martin:2014bb}
{Martin} R.~G.,  {Nixon} C.,  {Lubow} S.~H.,  {Armitage} P.~J.,  {Price} D.~J.,
   {Do{\u g}an} S.,   {King} A.,  2014b, \mn@doi [\apjl]
  {10.1088/2041-8205/792/2/L33}, \href
  {http://adsabs.harvard.edu/abs/2014ApJ...792L..33M} {792, L33}

\bibitem[\protect\citeauthoryear{{Martin}, {Lubow}, {Nixon}  \&
  {Armitage}}{{Martin} et~al.}{2016}]{Martin:2016qf}
{Martin} R.~G.,  {Lubow} S.~H.,  {Nixon} C.,   {Armitage} P.~J.,  2016, \mn@doi
  [\mnras] {10.1093/mnras/stw605}, \href
  {http://adsabs.harvard.edu/abs/2016MNRAS.458.4345M} {458, 4345}

\bibitem[\protect\citeauthoryear{{Marzari} \& {Nelson}}{{Marzari} \&
  {Nelson}}{2009}]{Marzari:2009of}
{Marzari} F.,  {Nelson} A.~F.,  2009, \mn@doi [\apj]
  {10.1088/0004-637X/705/2/1575}, \href
  {http://ukads.nottingham.ac.uk/abs/2009ApJ...705.1575M} {705, 1575}

\bibitem[\protect\citeauthoryear{{Min}, {Stolker}, {Dominik}  \&
  {Benisty}}{{Min} et~al.}{2017}]{Min:2017od}
{Min} M.,  {Stolker} T.,  {Dominik} C.,   {Benisty} M.,  2017, \mn@doi [\aap]
  {10.1051/0004-6361/201730949}, \href
  {http://ukads.nottingham.ac.uk/abs/2017A%26A...604L..10M} {604, L10}

\bibitem[\protect\citeauthoryear{{Nagasawa}, {Ida}  \& {Bessho}}{{Nagasawa}
  et~al.}{2008}]{Nagasawa:2008pj}
{Nagasawa} M.,  {Ida} S.,   {Bessho} T.,  2008, \mn@doi [\apj]
  {10.1086/529369}, \href
  {http://ukads.nottingham.ac.uk/abs/2008ApJ...678..498N} {678, 498}

\bibitem[\protect\citeauthoryear{{Nealon}, {Price}  \& {Nixon}}{{Nealon}
  et~al.}{2015}]{Nealon:2015fk}
{Nealon} R.,  {Price} D.~J.,   {Nixon} C.~J.,  2015, \mn@doi [\mnras]
  {10.1093/mnras/stv014}, \href
  {http://adsabs.harvard.edu/abs/2015MNRAS.448.1526N} {448, 1526}

\bibitem[\protect\citeauthoryear{{Nealon}, {Nixon}, {Price}  \&
  {King}}{{Nealon} et~al.}{2016}]{Nealon:2016lr}
{Nealon} R.,  {Nixon} C.,  {Price} D.~J.,   {King} A.,  2016, \mn@doi [\mnras]
  {10.1093/mnrasl/slv149}, \href
  {http://adsabs.harvard.edu/abs/2016MNRAS.455L..62N} {455, L62}

\bibitem[\protect\citeauthoryear{{Nixon}}{{Nixon}}{2012}]{nixon_2012}
{Nixon} C.~J.,  2012, \mn@doi [\mnras] {10.1111/j.1365-2966.2012.21072.x},
  \href {http://adsabs.harvard.edu/abs/2012MNRAS.423.2597N} {423, 2597}

\bibitem[\protect\citeauthoryear{{Nixon} \& {King}}{{Nixon} \&
  {King}}{2016}]{Nixon:2016nh}
{Nixon} C.,  {King} A.,  2016, {Warp Propagation in Astrophysical Discs}.
p.~45, \mn@doi{10.1007/978-3-319-19416-5_2}

\bibitem[\protect\citeauthoryear{{Nixon}, {King}  \& {Price}}{{Nixon}
  et~al.}{2013}]{nixon_2013}
{Nixon} C.,  {King} A.,   {Price} D.,  2013, \mn@doi [\mnras]
  {10.1093/mnras/stt1136}, \href
  {http://adsabs.harvard.edu/abs/2013MNRAS.434.1946N} {434, 1946}

\bibitem[\protect\citeauthoryear{{Nixon}, {King}  \& {Pringle}}{{Nixon}
  et~al.}{2018}]{Nixon:2018hf}
{Nixon} C.~J.,  {King} A.~R.,   {Pringle} J.~E.,  2018, \mn@doi [\mnras]
  {10.1093/mnras/sty593}, \href
  {http://adsabs.harvard.edu/abs/2018MNRAS.477.3273N} {477, 3273}

\bibitem[\protect\citeauthoryear{{Papaloizou} \& {Larwood}}{{Papaloizou} \&
  {Larwood}}{2000}]{Papaloizou:2000ks}
{Papaloizou} J.~C.~B.,  {Larwood} J.~D.,  2000, \mn@doi [\mnras]
  {10.1046/j.1365-8711.2000.03466.x}, \href
  {http://ukads.nottingham.ac.uk/abs/2000MNRAS.315..823P} {315, 823}

\bibitem[\protect\citeauthoryear{{Papaloizou} \& {Lin}}{{Papaloizou} \&
  {Lin}}{1995}]{Papaloizou:1995pn}
{Papaloizou} J.~C.~B.,  {Lin} D.~N.~C.,  1995, \mn@doi [\apj] {10.1086/175127},
  \href {http://adsabs.harvard.edu/abs/1995ApJ...438..841P} {438, 841}

\bibitem[\protect\citeauthoryear{{Papaloizou} \& {Pringle}}{{Papaloizou} \&
  {Pringle}}{1983}]{pap_pringle_1983}
{Papaloizou} J.~C.~B.,  {Pringle} J.~E.,  1983, \mnras, \href
  {http://adsabs.harvard.edu/abs/1983MNRAS.202.1181P} {202, 1181}

\bibitem[\protect\citeauthoryear{{Papaloizou} \& {Terquem}}{{Papaloizou} \&
  {Terquem}}{1995}]{Papaloizou:1995fk}
{Papaloizou} J.~C.~B.,  {Terquem} C.,  1995, \mn@doi [\mnras]
  {10.1093/mnras/274.4.987}, \href
  {http://adsabs.harvard.edu/abs/1995MNRAS.274..987P} {274, 987}

\bibitem[\protect\citeauthoryear{{Papaloizou}, {Nelson}  \&
  {Masset}}{{Papaloizou} et~al.}{2001}]{Papaloizou:2001ng}
{Papaloizou} J.~C.~B.,  {Nelson} R.~P.,   {Masset} F.,  2001, \mn@doi [\aap]
  {10.1051/0004-6361:20000011}, \href
  {http://ukads.nottingham.ac.uk/abs/2001A%26A...366..263P} {366, 263}

\bibitem[\protect\citeauthoryear{{P{\'e}rez}, {Isella}, {Carpenter}  \&
  {Chandler}}{{P{\'e}rez} et~al.}{2014}]{Perez:2014bq}
{P{\'e}rez} L.~M.,  {Isella} A.,  {Carpenter} J.~M.,   {Chandler} C.~J.,  2014,
  \mn@doi [\apjl] {10.1088/2041-8205/783/1/L13}, \href
  {http://adsabs.harvard.edu/abs/2014ApJ...783L..13P} {783, L13}

\bibitem[\protect\citeauthoryear{{Pinilla}, {Birnstiel}  \& {Walsh}}{{Pinilla}
  et~al.}{2015}]{Pinilla:2015ci}
{Pinilla} P.,  {Birnstiel} T.,   {Walsh} C.,  2015, \mn@doi [\aap]
  {10.1051/0004-6361/201425539}, \href
  {http://ukads.nottingham.ac.uk/abs/2015A%26A...580A.105P} {580, A105}

\bibitem[\protect\citeauthoryear{{Pinte}, {Dent}, {M{\'e}nard}, {Hales},
  {Hill}, {Cortes}  \& {de Gregorio-Monsalvo}}{{Pinte}
  et~al.}{2016}]{Pinte:2016oa}
{Pinte} C.,  {Dent} W.~R.~F.,  {M{\'e}nard} F.,  {Hales} A.,  {Hill} T.,
  {Cortes} P.,   {de Gregorio-Monsalvo} I.,  2016, \mn@doi [\apj]
  {10.3847/0004-637X/816/1/25}, \href
  {http://ukads.nottingham.ac.uk/abs/2016ApJ...816...25P} {816, 25}

\bibitem[\protect\citeauthoryear{{Pontoppidan}, {Blake}, {van Dishoeck},
  {Smette}, {Ireland}  \& {Brown}}{{Pontoppidan}
  et~al.}{2008}]{Pontoppidan:2008ge}
{Pontoppidan} K.~M.,  {Blake} G.~A.,  {van Dishoeck} E.~F.,  {Smette} A.,
  {Ireland} M.~J.,   {Brown} J.,  2008, \mn@doi [\apj] {10.1086/590400}, \href
  {http://adsabs.harvard.edu/abs/2008ApJ...684.1323P} {684, 1323}

\bibitem[\protect\citeauthoryear{{Poteet} et~al.,}{{Poteet}
  et~al.}{2018}]{Poteet:2018be}
{Poteet} C.~A.,  et~al., 2018, preprint, \href
  {http://adsabs.harvard.edu/abs/2018arXiv180501926P} {} (\mn@eprint {arXiv}
  {1805.01926})

\bibitem[\protect\citeauthoryear{{Price}}{{Price}}{2007}]{Price:2007kx}
{Price} D.~J.,  2007, \mn@doi [\pasa] {10.1071/AS07022}, \href
  {http://adsabs.harvard.edu/abs/2007PASA...24..159P} {24, 159}

\bibitem[\protect\citeauthoryear{{Price} et~al.,}{{Price}
  et~al.}{2017}]{Phantom}
{Price} D.~J.,  et~al., 2017, preprint, \href
  {http://adsabs.harvard.edu/abs/2017arXiv170203930P} {} (\mn@eprint {arXiv}
  {1702.03930})

\bibitem[\protect\citeauthoryear{{Price} et~al.,}{{Price}
  et~al.}{2018}]{Price:2018pf}
{Price} D.~J.,  et~al., 2018, \mn@doi [\mnras] {10.1093/mnras/sty647}, \href
  {http://ukads.nottingham.ac.uk/abs/2018MNRAS.477.1270P} {477, 1270}

\bibitem[\protect\citeauthoryear{{Pringle}}{{Pringle}}{1981}]{Pringle:1981fo}
{Pringle} J.~E.,  1981, \mn@doi [\araa] {10.1146/annurev.aa.19.090181.001033},
  \href {http://adsabs.harvard.edu/abs/1981ARA%26A..19..137P} {19, 137}

\bibitem[\protect\citeauthoryear{{Pringle}}{{Pringle}}{1996}]{pringle_1996}
{Pringle} J.~E.,  1996, \mnras, \href
  {http://adsabs.harvard.edu/abs/1996MNRAS.281..357P} {281, 357}

\bibitem[\protect\citeauthoryear{{Qi} et~al.,}{{Qi} et~al.}{2004}]{Qi:2004bw}
{Qi} C.,  et~al., 2004, \mn@doi [\apjl] {10.1086/421063}, \href
  {http://adsabs.harvard.edu/abs/2004ApJ...616L..11Q} {616, L11}

\bibitem[\protect\citeauthoryear{{Quillen}}{{Quillen}}{2006}]{Quillen:2006bv}
{Quillen} A.~C.,  2006, \mn@doi [\apj] {10.1086/500165}, \href
  {http://adsabs.harvard.edu/abs/2006ApJ...640.1078Q} {640, 1078}

\bibitem[\protect\citeauthoryear{{Ragusa}, {Dipierro}, {Lodato}, {Laibe}  \&
  {Price}}{{Ragusa} et~al.}{2017}]{Ragusa:2017nv}
{Ragusa} E.,  {Dipierro} G.,  {Lodato} G.,  {Laibe} G.,   {Price} D.~J.,  2017,
  \mn@doi [\mnras] {10.1093/mnras/stw2456}, \href
  {http://ukads.nottingham.ac.uk/abs/2017MNRAS.464.1449R} {464, 1449}

\bibitem[\protect\citeauthoryear{{Rein}}{{Rein}}{2012}]{Rein:2012so}
{Rein} H.,  2012, \mn@doi [\mnras] {10.1111/j.1365-2966.2012.20869.x}, \href
  {http://ukads.nottingham.ac.uk/abs/2012MNRAS.422.3611R} {422, 3611}

\bibitem[\protect\citeauthoryear{{Ru{\'{\i}}z-Rodr{\'{\i}}guez}, {Ireland},
  {Cieza}  \& {Kraus}}{{Ru{\'{\i}}z-Rodr{\'{\i}}guez}
  et~al.}{2016}]{Ruiz:2016ne}
{Ru{\'{\i}}z-Rodr{\'{\i}}guez} D.,  {Ireland} M.,  {Cieza} L.,   {Kraus} A.,
  2016, \mn@doi [\mnras] {10.1093/mnras/stw2297}, \href
  {http://adsabs.harvard.edu/abs/2016MNRAS.463.3829R} {463, 3829}

\bibitem[\protect\citeauthoryear{Shakura \& Sunyaev}{Shakura \&
  Sunyaev}{1973}]{shakura_sunyaev}
Shakura N.,  Sunyaev R.,  {1973}, \aap, {24}, 337

\bibitem[\protect\citeauthoryear{{Stolker} et~al.,}{{Stolker}
  et~al.}{2016}]{Stolker:2016ck}
{Stolker} T.,  et~al., 2016, \mn@doi [\aap] {10.1051/0004-6361/201528039},
  \href {http://ukads.nottingham.ac.uk/abs/2016A%26A...595A.113S} {595, A113}

\bibitem[\protect\citeauthoryear{{Stolker} et~al.,}{{Stolker}
  et~al.}{2017}]{Stolker:2017of}
{Stolker} T.,  et~al., 2017, preprint, \href
  {http://ukads.nottingham.ac.uk/abs/2017arXiv171002532S} {} (\mn@eprint
  {arXiv} {1710.02532})

\bibitem[\protect\citeauthoryear{{Tanaka} \& {Ward}}{{Tanaka} \&
  {Ward}}{2004}]{Tanaka:2004od}
{Tanaka} H.,  {Ward} W.~R.,  2004, \mn@doi [\apj] {10.1086/380992}, \href
  {http://ukads.nottingham.ac.uk/abs/2004ApJ...602..388T} {602, 388}

\bibitem[\protect\citeauthoryear{{Tanaka}, {Takeuchi}  \& {Ward}}{{Tanaka}
  et~al.}{2002}]{Tanaka:2002of}
{Tanaka} H.,  {Takeuchi} T.,   {Ward} W.~R.,  2002, \mn@doi [\apj]
  {10.1086/324713}, \href
  {http://ukads.nottingham.ac.uk/abs/2002ApJ...565.1257T} {565, 1257}

\bibitem[\protect\citeauthoryear{{Teyssandier} \& {Ogilvie}}{{Teyssandier} \&
  {Ogilvie}}{2017}]{Teyssandier:2017uh}
{Teyssandier} J.,  {Ogilvie} G.~I.,  2017, \mn@doi [\mnras]
  {10.1093/mnras/stx426}, \href
  {http://ukads.nottingham.ac.uk/abs/2017MNRAS.467.4577T} {467, 4577}

\bibitem[\protect\citeauthoryear{{Teyssandier}, {Terquem}  \&
  {Papaloizou}}{{Teyssandier} et~al.}{2013}]{Teyssandier:2013og}
{Teyssandier} J.,  {Terquem} C.,   {Papaloizou} J.~C.~B.,  2013, \mn@doi
  [\mnras] {10.1093/mnras/sts064}, \href
  {http://ukads.nottingham.ac.uk/abs/2013MNRAS.428..658T} {428, 658}

\bibitem[\protect\citeauthoryear{{Thommes} \& {Lissauer}}{{Thommes} \&
  {Lissauer}}{2003}]{Thommes:2003hb}
{Thommes} E.~W.,  {Lissauer} J.~J.,  2003, \mn@doi [\apj] {10.1086/378317},
  \href {http://ukads.nottingham.ac.uk/abs/2003ApJ...597..566T} {597, 566}

\bibitem[\protect\citeauthoryear{{Triaud} et~al.,}{{Triaud}
  et~al.}{2010}]{Triaud:2010vj}
{Triaud} A.~H.~M.~J.,  et~al., 2010, \mn@doi [\aap]
  {10.1051/0004-6361/201014525}, \href
  {http://ukads.nottingham.ac.uk/abs/2010A%26A...524A..25T} {524, A25}

\bibitem[\protect\citeauthoryear{{Tsukagoshi} et~al.,}{{Tsukagoshi}
  et~al.}{2016}]{Tsukagoshi:2016hf}
{Tsukagoshi} T.,  et~al., 2016, \mn@doi [\apjl] {10.3847/2041-8205/829/2/L35},
  \href {http://ukads.nottingham.ac.uk/abs/2016ApJ...829L..35T} {829, L35}

\bibitem[\protect\citeauthoryear{{Uyama}, {Tanigawa}, {Hashimoto}, {Tamura},
  {Aoyama}, {Brandt}  \& {Ishizuka}}{{Uyama} et~al.}{2017}]{Uyama:2016ja}
{Uyama} T.,  {Tanigawa} T.,  {Hashimoto} J.,  {Tamura} M.,  {Aoyama} Y.,
  {Brandt} T.~D.,   {Ishizuka} M.,  2017, \mn@doi [\aj]
  {10.3847/1538-3881/aa816a}, \href
  {http://ukads.nottingham.ac.uk/abs/2017AJ....154...90U} {154, 90}

\bibitem[\protect\citeauthoryear{{Walsh}, {Daley}, {Facchini}  \&
  {Juhasz}}{{Walsh} et~al.}{2017}]{Walsh:2017ic}
{Walsh} C.,  {Daley} C.,  {Facchini} S.,   {Juhasz} A.,  2017, preprint, \href
  {http://ukads.nottingham.ac.uk/abs/2017arXiv171000703W} {} (\mn@eprint
  {arXiv} {1710.00703})

\bibitem[\protect\citeauthoryear{{Williams} \& {Cieza}}{{Williams} \&
  {Cieza}}{2011}]{Williams:2011he}
{Williams} J.~P.,  {Cieza} L.~A.,  2011, \mn@doi [\araa]
  {10.1146/annurev-astro-081710-102548}, \href
  {http://ukads.nottingham.ac.uk/abs/2011ARA%26A..49...67W} {49, 67}

\bibitem[\protect\citeauthoryear{{Xiang-Gruess} \& {Papaloizou}}{{Xiang-Gruess}
  \& {Papaloizou}}{2013}]{Xiang-Gruess:2013fg}
{Xiang-Gruess} M.,  {Papaloizou} J.~C.~B.,  2013, \mn@doi [\mnras]
  {10.1093/mnras/stt254}, \href
  {http://ukads.nottingham.ac.uk/abs/2013MNRAS.431.1320X} {431, 1320}

\bibitem[\protect\citeauthoryear{{de Val-Borro} et~al.,}{{de Val-Borro}
  et~al.}{2006}]{deVal-Borro:2006fd}
{de Val-Borro} M.,  et~al., 2006, \mn@doi [\mnras]
  {10.1111/j.1365-2966.2006.10488.x}, \href
  {http://ukads.nottingham.ac.uk/abs/2006MNRAS.370..529D} {370, 529}

\bibitem[\protect\citeauthoryear{{van der Marel}, {van Dishoeck}, {Bruderer},
  {P{\'e}rez}  \& {Isella}}{{van der Marel} et~al.}{2015}]{vanderMarel:2015ne}
{van der Marel} N.,  {van Dishoeck} E.~F.,  {Bruderer} S.,  {P{\'e}rez} L.,
  {Isella} A.,  2015, \mn@doi [\aap] {10.1051/0004-6361/201525658}, \href
  {http://adsabs.harvard.edu/abs/2015A%26A...579A.106V} {579, A106}

\bibitem[\protect\citeauthoryear{{van der Marel} et~al.,}{{van der Marel}
  et~al.}{2018}]{vanderMarel:2018ve}
{van der Marel} N.,  et~al., 2018, \mn@doi [\apj] {10.3847/1538-4357/aaaa6b},
  \href {http://adsabs.harvard.edu/abs/2018ApJ...854..177V} {854, 177}

\makeatother
\end{thebibliography}

\appendix
\section{Resolution study}
\label{section:racc_study}
The offset between the left and right panels of Fig.~\ref{fig:radial_migration} and \ref{fig:inclination_damping} demonstrate that the acceleration experienced by the planet is smaller than the rate at which it actually migrates. Previous work using SPH by \citet{Ayliffe:2010ow} indicates that the migration rate strongly depends on the accretion radius chosen \citep[but this is also true for grid simulations, e.g.][]{deVal-Borro:2006fd}. We thus conduct a convergence study of the accretion radius used, taking the simulation at $i=1.62^{\circ}$, $N=1.0 \times 10^6$ and $R_{\rm acc}~=~0.25R_{\rm Hill}$ as our reference simulation. This simulation is conducted two additional times, using the smaller and larger accretion radii $R_{\rm acc}~=~0.125, 0.40 R_{\rm Hill}$. These numerical tests are named G2 and G3 in Table~\ref{tab:sims_summary}, respectively.

\begin{figure}
	\includegraphics[width=\columnwidth]{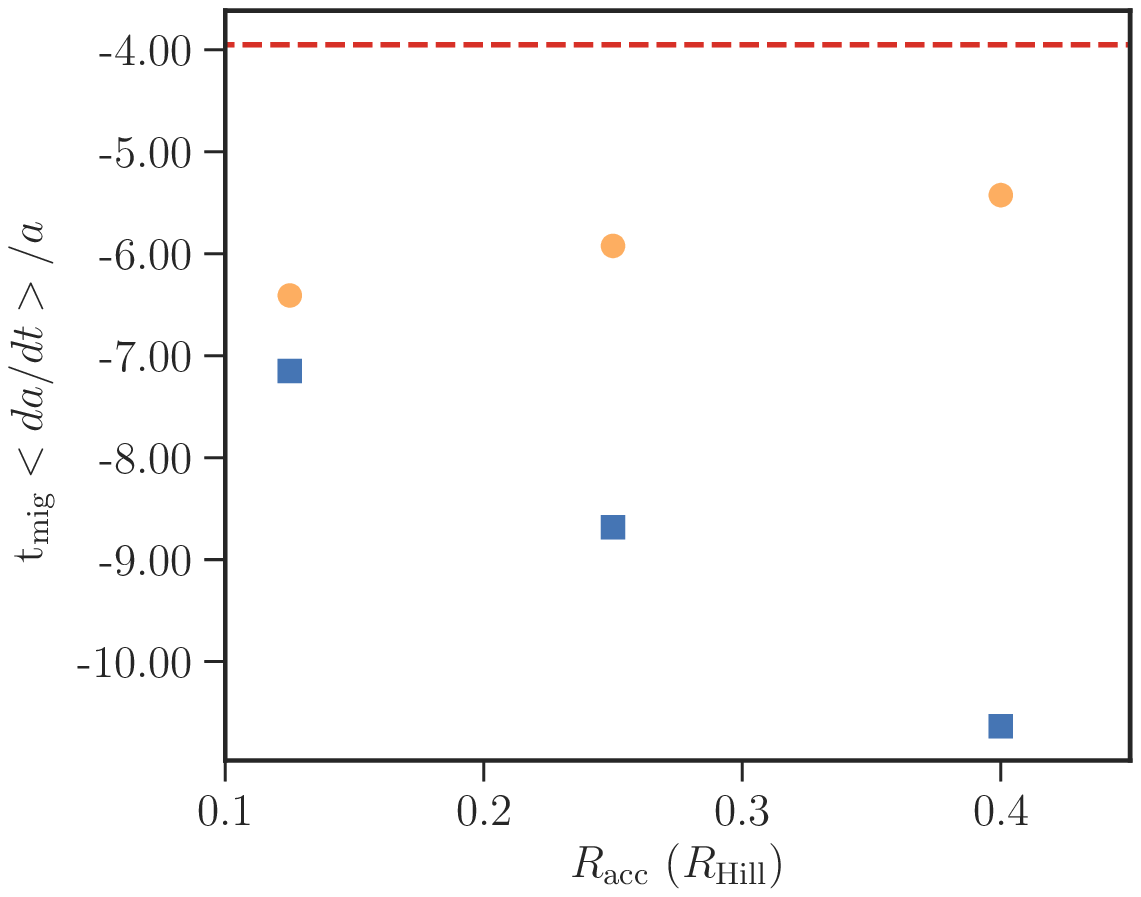}
	\includegraphics[width=\columnwidth]{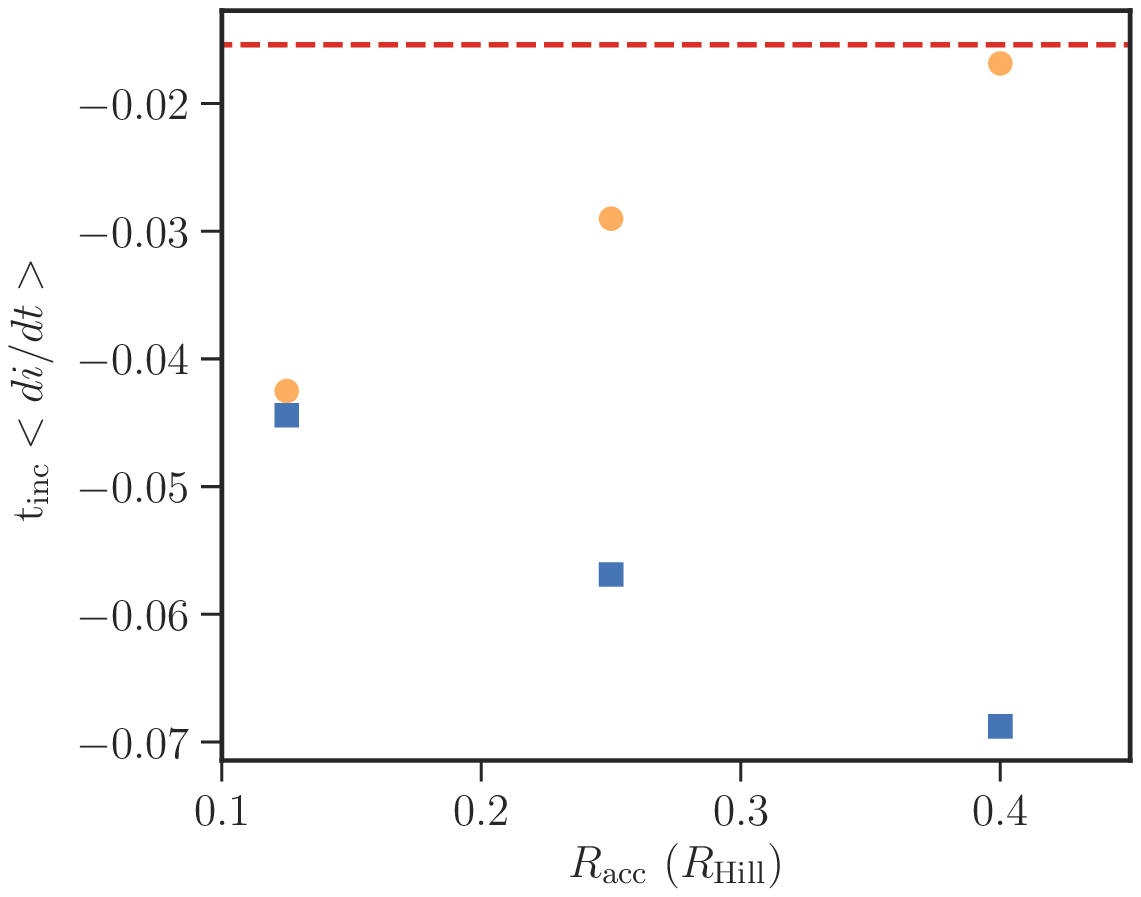}
    \caption{Radial migration (upper panel) and inclination damping rate (lower panel) calculated from both the accelerations experienced by the planet (orange circles) and the actual migration (blue squares) for three accretion radii. The inviscid, analytical approximations from Equation~\ref{equation:tanaka_2002} is shown in the upper panel and from Equation~\ref{equation:tanaka_2004} in the lower panel with the dashed red line. Decreasing the accretion radius leads to better resolution of the flow immediately around the planet, improving the accuracy of the planets migration while altering the surface density profile immediately around the planet. \label{fig:racc_convergence}}
\end{figure}

Fig.~\ref{fig:racc_convergence} shows the radial migration rate and inclination damping rates (as calculated in Sections~\ref{subsection:radial_migration}~and~\ref{subsection:inclination_damping}), comparing the three different accretion radii for the two different methods of calculating the damping rates. In agreement with \citet{Ayliffe:2010ow}, when measuring the damping from the motion of the planet (blue squares) we find closer agreement to the analytical predictions \citep[shown with the red lines][]{Tanaka:2002of,Tanaka:2004od} with decreasing accretion radius. As explored at length in \citet{Ayliffe:2010ow}, with smaller accretion radii there is a more accurate description of the flow around the planet including the material that moves into the Hill sphere, circulates and exits (e.g. their fig. 11). As this material is able to exchange angular momentum with the planet, resolving this region more accurately leads to better accordance between the two methods of calculating the damping rates. Even with our largest accretion radius of 0.4$R_{\rm Hill}$, the migration rate is still within a factor of three of the expectation for the radial migration and a five for the inclination damping time-scale.

Fig.~\ref{fig:racc_convergence} also shows the migration and inclination damping rates measured from the accelerations (orange circles). With decreasing accretion radius we find that the migration rate speeds up for both damping time-scales, moving away from the analytical expectation. This trend is likely due to the different accretion radii cutting out successively smaller portions of the accretion disc, altering the surface density profile and hence the accelerations measured from the surrounding gas particles \citep{Ayliffe:2010ow}. Although decreasing the accretion radius moves the expected migration rate away from the analytical prediction, the inclusion of viscosity in our simulations means that we do not necessarily expect to match the inviscid predictions from \citet{Tanaka:2002of} and \citet{Tanaka:2004od}. As a smaller accretion radius is more computationally expensive and we are modelling behaviour over hundreds of orbits, we have used $R_{\rm acc} = 0.25, 0.40 R_{\rm Hill}$ for the simulations presented here.

\bsp	
\label{lastpage}
\end{document}